\documentclass[11pt,a4paper]{article}

\usepackage{hyperref}

\usepackage[fleqn]{amsmath}
\usepackage{amsthm,amssymb,epsfig,latexsym,euscript}

\usepackage{pgf}
\usepackage{tikz}

\newtheorem{thm}{Theorem}[section]
\newtheorem{prop}{Proposition}[section]
\newtheorem{lemma}{Lemma}[section]
\newtheorem{cor}{Corollary}[section]

{\theoremstyle{remark}
\newtheorem{rem}{Remark}[section]}
{\theoremstyle{remark}
}
\def\Proof{\medskip\noindent {\it Proof --- \ }}

\let\qed=\cqfd

\makeatletter
\@addtoreset{equation}{section}
\makeatother

\newcommand{\tr}{\operatorname{tr}}

\newcommand{\bra}[1]{\langle\,#1\,|}
\newcommand{\ket}[1]{|\,#1\,\rangle}

\newcommand{\moy}[1]{\langle\,#1\,\rangle}

\def\eps{\epsilon}

\newcommand{\e}{\mathsf{e}}

\begin{document}

\begin{flushright}
LPENSL-TH-11/12
\end{flushright}

\vspace{24pt}

\begin{center}
\begin{LARGE}
{\bf Algebraic Bethe Ansatz approach to form factors and correlation functions
of the cyclic eight-vertex solid-on-solid model}
\end{LARGE}

\vspace{50pt}

\begin{large}
{\bf D.~Levy-Bencheton}\footnote{damien.levybencheton@ens-lyon.fr}
{\bf and V.~Terras}\footnote{veronique.terras@ens-lyon.fr}
\end{large}

\vspace{.5cm}
 
{Laboratoire de Physique, ENS Lyon \& CNRS UMR 5672,\\
Universit\'e de Lyon, France}
\vspace{2cm}

\today

\end{center}

\vspace{1cm}

\begin{abstract}
We consider the problem of the exact computation of the correlation functions of the eight-vertex solid-on-solid model by means of the algebraic Bethe Ansatz.
We compute the scalar product between a Bethe eigenstate and an arbitrary state of Bethe type and show that, in the cyclic case, it can be formulated as a single determinant of usual functions.
It allows us to obtain determinant representations for finite-size form factors.
By summing up over the form factors, we also give a multiple integral representation for a generating function of the two-point function.

\end{abstract}

\vspace{1cm}

\section{Introduction}
\label{sec-intro}

In the domain of integrable systems, after the successful determination of the spectrum and the eigenstates of the Hamiltonian \cite{Bet31,FadST79,Tha81,Bax82L,Gau83L,Mat93L,Fad96}, a crucial problem is the computation of the form factors and correlation functions which are essential objects for the description of the full dynamical properties of the models.
For truly interacting integrable models (i.e. for models which are not equivalent to free fermions), this 
problem represents a real challenge and has been intensively studied during the last decades (see e.g.  \cite{KarW78,CarM90,FriMS93,Zam91,Smi92L,BogIK93L,JimMMN92,JimM95L,JimM96,LukZ97,Luk99,LukZ01,LukT03,KitMT99,KitMT00,KitMST02a,KitMST05a,KitKMST09b,BooJMST06a,BooJMST06b,BooJMST06c,BooJMST07,BooG09,BooJMS09,JimMS09,JimMS11,JimMS11a,BouKM98,SatST04,CauM05}).
Several methods have been developed so far, and important progresses have been made recently, in particular for the archetypal integrable models such as the isotropic (XXX) and partially anisotropic (XXZ) Heisenberg chains, or the one-dimensional quantum Bose gas.

Among these methods, one \cite{KitMT99,KitMT00}, based on algebraic Bethe Ansatz (ABA) \cite{FadST79,Fad96}, has led to important recent developments \cite{KitMST02a,GohKS04,GohKS05,KitMST05a,KitMST05b,Sak07,KitKMST07,KitKMST09b,KozT11,Koz11,KitKMST11b,KitKMST12}.
It relies on the solution of the quantum inverse problem \cite{KitMT99,MaiT00, GohK00} and on a compact and explicit expression, in the form of a determinant of usual functions, for the scalar products between a Bethe eigenstate and an arbitrary state of Bethe type \cite{Sla89}.
The combination of these two results leads to simple determinant representations for the form factors (i.e. for the matrix elements of local operators in the basis of eigenstates of the transfer matrix) in finite volume \cite{KitMT99}.
Although the exact representations which are obtained for the physical correlation functions (and in particular for the two-point functions) have a less simple form \cite{KitMST02a,KitMST04c,KitMST05a,KitMST05b,KitKMST07}, it is nevertheless possible to obtain quantitative results for these quantities as well, for instance by summation over the corresponding form factors series.
This can be achieved either by numerical methods relying on the efficient, explicit aforementioned determinant representation for the form factors \cite{BieKM03,CauHM05,CauM05} or, in the asymptotic regime, from the analytic study of the series \cite{KitKMST09b,KozT11,Koz11,KitKMST11b,KitKMST12}.

However, until now, this method has been essentially developed in the relatively simple case of the periodic XXZ spin-1/2 Heisenberg chain or some of its variants (open chain \cite{KitKMNST07,KitKMNST08}, higher spins \cite{Kit01,GohSS10,Deg12}
\ldots), or in the even simpler case of the Bose gas model \cite{KitKMST07,KozT11,KitKMST12}.
In fact, for more complicated models, the solution of one of the two aforementioned basic ingredients (resolution of the inverse problem and determinant representation for the scalar products of Bethe states) is often missing.

In particular, a real challenge would be to adapt this method so as to find explicit and manageable expressions for the correlation functions of the completely anisotropic (XYZ) Heisenberg chain.
The latter is a natural generalization of the XXZ Heisenberg chain, and is related to the eight-vertex model of statistical physics (which in its turn is a natural generalization of the six-vertex model related to the XXZ chain).
It was solved for the first time by Baxter \cite{Bax71a,Bax71b,Bax72,Bax73a} (see also \cite{JohKM73} for the computation of the correlation length, and \cite{FadT79,FelV96b} for the ABA formulation).
In fact, it is worth mentioning here that,  due to the non-conservation of the spin, the XYZ (or eight-vertex) model is not {\em directly} solvable by Bethe Ansatz. Nevertheless, Baxter managed to construct the eigenstates of this model by relating them (via the so-called Vertex-IRF transformation) to the Bethe eigenstates of another model of solid-on-solid (SOS) type.
The latter, which is often called eight-vertex SOS model (8VSOS), is a $L$-state IRF (interaction-round-a-face) model on a square lattice and is solvable by Bethe Ansatz \cite{Bax73a,FelV96b}.
However, due to the need of the Vertex-IRF transformation to come back to the XYZ (or eight-vertex) model, the resolution process is therefore
much more complicated than in the XXZ case, and the exact computation of the correlation functions is in this case still a widely open problem. In particular, through the aforementioned method based on ABA, no significant results in this direction have been obtained so far (see nevertheless, by means of another approach using $q$-vertex operators \cite{JimM95L}, some first results based on  those of \cite{LukP96} in \cite{LasP98,Las02}, as well as an attempt of a direct approach in \cite{Shi04}).

Actually, the combinatorial complexity issued from the use of the Vertex-IRF transformation is not the only problem for the computation of the XYZ correlation functions. In fact, even in the simpler case of the related SOS models \cite{AndBF84,KunY88,PeaS89,PeaB90}, the ABA approach to correlation functions
presents some difficulties, in particular concerning the obtention of a compact and manageable formula for the scalar products of states.
Let us recall at this point that, in the XXZ case, the scalar product between a Bethe eigenstate and an arbitrary state of Bethe type can be written as a determinant of usual functions \cite{Sla89}, and that the ABA approach to correlation functions relies strongly on such a representation \cite{KitMT99}.
The latter originates (see for instance the derivation of \cite{KitMT99}) from the explicit  determinant representation of the partition function of the six-vertex model with special boundary conditions, the so-called domain wall boundary conditions \cite{Kor82,Ize87}.
The main problem in the 8VSOS case is that  the analogous partition function with domain wall boundary conditions does not seem to admit a representation in the form of a single determinant \cite{Ros09,PakRS08} (see also \cite{Gal12b}). 
This is due to the fact that the $R$-matrix of the model depends (compared to the simpler six-vertex $R$-matrix) on an extra dynamical parameter related to the fluctuation variable (height) of the model.
This $R$-matrix hence satisfies the dynamical version of the Yang-Baxter equation \cite{Fel95,FelV96a,FelV96b} instead of the usual one. As a result, the dynamical parameter is subject to some shifts, which prevents one from reducing the sum of determinants obtained in \cite{Ros09} for the 8VSOS partition function to a unique one as in \cite{Ize87} (see nevertheless \cite{FilK10,Fil11} for a model with a reflecting end, for which the partition function indeed reduces to a unique determinant of Izergin type).
Hence, a mere generalization of the process described in \cite{KitMT99} seems to be not so easy.

This is the problem that we tackle in this article, namely to set the bases of the ABA approach to correlation functions in the dynamical case, i.e. in the case of the periodic 8VSOS model.
Building on Rosengren's representation \cite{Ros09} of the partition function with domain wall boundary conditions as a sum of determinants, we obtain a similar representation (i.e. as a sum of determinants) for the partial scalar product, evaluated at a fixed value of the height $s$, of a Bethe eigenstate with an arbitrary state.
Hence, the fact that we do not {\em a priori} obtain a single determinant may be a problem for the computation of correlation functions.
However we show that, in the cyclic case (i.e. when the parameter $\eta$ of the model is rational and the space of states is finite-dimensional), the true scalar product of Bethe states, which is obtained by summing up over all values of the dynamical parameter, can in fact be represented as a unique determinant. In the same way, the finite-size form factors (i.e. the matrix elements, in the basis of eigenstates of the finite-size transfer matrix, of local operators labeling the difference of height between two neighboring vertices) can be written as a unique determinant which has a similar form as in XXZ.
This opens the way to the computation of correlation functions, for instance by summation over the form factor series.
In particular, we define a generating function of the two-point function and show that, in the cyclic dynamical model in finite volume, the latter admits a multiple integral representation very similar to the {\em master equation} representation that was obtained in \cite{KitMST05a} and used in \cite{KitKMST09b} to derive the long-distance asymptotic behavior for the two-point function in the XXZ case.

The content of the article is the following.
In Section~\ref{sec-SOS}, we define the 8VSOS model and recall the main steps of its algebraic Bethe Ansatz resolution.
In Section~\ref{sec-sc-pdt}, we compute the scalar product between a Bethe eigenstate and an arbitrary state of Bethe type.
In Section~\ref{sec-inv}, we solve the quantum inverse problem in the dynamical case, namely we express local spin operators in terms of elements of the dynamical monodromy matrix.
In Section~\ref{sec-ff}, we compute the finite-size form factors.
In Section~\ref{sec-master}, we formally sum up the form factor series and obtain a multiple integral representation (master equation representation) for a generating function of the two-point function of the (finite-size) cyclic dynamical model.
Details and technicalities are gathered in a set of appendices.

\section{The 8VSOS model and algebraic Bethe Ansatz}
\label{sec-SOS}

Let us consider a two-dimensional square lattice with $N\times M$ elementary square cells (or faces), with periodic (toroidal) boundary conditions.
A height $s$ is attached to each vertex (site) of the lattice, so that heights on adjacent sites are restricted to differ by $\pm 1$, and a complex parameter $u_i$ (resp. $\xi_j$) is attached to each column $i$ (resp. line $j$) of cells.
To each height configuration $(s_1,s_2,s_3,s_4)$ (with $|s_1-s_2|=|s_2-s_3|=|s_3-s_4|=|s_4-s_1|=1$) around an elementary face of the lattice is associated a statistical weight $W\binom{s_1\ s_2}{s_4\ s_3}$, which can be understood as a matrix element of some matrix $R$ depending on the difference of the two corresponding parameters:

\begin{pgfpicture}{0cm}{0cm}{2cm}{2cm}
 \pgfputat{\pgfxy(0.35,1)}{\pgfbox[left,center]{
 ${R}(u_i-\xi_j; s)^{\epsilon_i,\epsilon_j}_{\epsilon'_i,\epsilon'_j}=$}}
 \pgfline{\pgfxy(6.5,0.5)}{\pgfxy(6.5,1.5)}
 \pgfline{\pgfxy(5.5,0.5)}{\pgfxy(5.5,1.5)}
 \pgfline{\pgfxy(5.5,0.5)}{\pgfxy(6.5,0.5)}
 \pgfline{\pgfxy(5.5,1.5)}{\pgfxy(6.5,1.5)}
 
 \pgfsetendarrow{\pgfarrowto}
 \pgfsetdash{{3pt}{3pt}}{0pt}
 \pgfline{\pgfxy(6.7,1)}{\pgfxy(5.3,1)}
 \pgfline{\pgfxy(6,1.7)}{\pgfxy(6,0.3)}
 \pgfstroke

 \pgfputat{\pgfxy(5.4,1.6)}{\pgfbox[right,center]{$s$}}
 \pgfputat{\pgfxy(6.6,1.6)}{\pgfbox[left,center]{$s+\epsilon'_i$}}
 \pgfputat{\pgfxy(6,0)}{\pgfbox[center,center]{$u_i$}}
 \pgfputat{\pgfxy(5.45,0.1)}{\pgfbox[right,bottom]{$s+\epsilon_j$}}
 \pgfputat{\pgfxy(6.6,0.2)}{\pgfbox[left,bottom]{$s+\epsilon_i+\epsilon_j$}}
 \pgfputat{\pgfxy(6.6,-0.2)}{\pgfbox[left,bottom]{$=s+\epsilon'_i+\epsilon'_j$}}
 \pgfputat{\pgfxy(5,1)}{\pgfbox[center,center]{$\xi_j$}}
 
\pgfputat{\pgfxy(9,1)}{\pgfbox[left,center]{
 $\equiv{W}\binom{s\qquad s+\epsilon'_i}{s+\epsilon_j\ s+\eps_i+\epsilon_j}.$}} 
\end{pgfpicture}

\smallskip
\noindent
Here $\epsilon_i,\epsilon'_i,\epsilon_j,\epsilon'_j\in\{+1,-1\}$, and the matrix $R(u;s)\in\mathrm{End}(V\otimes V)$, where $V\sim\mathbb{C}^2$ is a two-dimensional vector space with basis $(\e_+,\e_-)$, is such that
\begin{equation}
  R(u;s)\, (\e_{\epsilon'_1}\otimes \e_{\epsilon'_2})
  =\sum_{\substack{\epsilon_1,\epsilon_2\\ \epsilon_1+\epsilon_2=\epsilon'_1+\epsilon'_2}}R(u;s)^{\epsilon_1,\epsilon_2}_{\epsilon'_1,\epsilon'_2}\,
  (\e_{\epsilon_1}\otimes \e_{\epsilon_2}).
\end{equation}  

The $R$-matrix of the SOS model admits only six non-zero elements, which can be parameterized as follows:
\begin{equation}\label{R-mat}
  R(u;s)=
  \begin{pmatrix} 1 & 0 & 0 & 0 \\
                              0 & b(u;s) & c(u;s) & 0 \\
                              0 & \bar{c}(u;s) & \bar{b}(u;s) & 0 \\
                              0 & 0 & 0 & 1 
  \end{pmatrix}
\end{equation}
with
\begin{align}
 &b(u;s)=\frac{[s+1] \,  [u]}{[s] \, [u+1]},           \quad   &  &c(u;s)=\frac{[s+u] \,  [1]}{[s] \, [u+1]} , \label{bc1}\\
 &\bar{b}(u;s)=\frac{[s-1] \,  [u]}{[s] \, [u+1]}=b(u;-s) , \quad  &  
 &\bar{c}(u;s)=\frac{[s-u] \,  [1]}{[s] \, [u+1]}=c(u;-s) . \label{bc2}
\end{align} 
The parameters $u$ and $s$ are respectively called spectral and dynamical parameters.
The function $u\mapsto [u]$ is an entire, odd and quasi-periodic function of quasi-periods $1/\eta$ and $\tau/\eta$, where $\eta$ and $\tau$ are two parameters such that $\Im\tau>0$ (see Appendix~\ref{ap-theta} for the precise definition of this function).
Graphically, the above non-zero matrix elements correspond to the six following statistical  weights:

\begin{tikzpicture}
    \draw(0,5.5) ;
    \draw(0,-2) ;   

    \draw (0,3.5) -- (0,4.5) node[above left]{$s$} ; 
    \draw (0,4.5) -- (1,4.5) node[above right]{$s+1$} ;
    \draw (1,4.5) -- (1,3.5) node[below right]{$s+2$} ;
    \draw (1,3.5) -- (0,3.5) node[below left]{$s+1$} ;
    \draw (0.5,4.5) node[above]{$+$} ;
    \draw (0.5,3.5) node[below]{$+$} ;
    \draw (0,4) node[left]{$+$} ;
    \draw (1,4) node[right]{$+$} ;
    
    \draw (0.5,2) node[above]{${a}(u;s)=1$} ;

    \draw (0,0) -- (0,1) node[above left]{$s$} ; 
    \draw (0,1) -- (1,1) node[above right]{$s-1$} ;
    \draw (1,1) -- (1,0) node[below right]{$s-2$} ;
    \draw (1,0) -- (0,0) node[below left]{$s-1$} ;
    \draw (0.5,1) node[above]{$-$} ;
    \draw (0.5,0) node[below]{$-$} ;
    \draw (0,0.5) node[left]{$-$} ;
    \draw (1,0.5) node[right]{$-$} ;  
    
    \draw (0.5,-1.5) node[above]{$\bar{a}(u;s)=1$} ;

    \draw (4,3.5) -- (4,4.5) node[above left]{$s$} ; 
    \draw (4,4.5) -- (5,4.5) node[above right]{$s+1$} ;
    \draw (5,4.5) -- (5,3.5) node[below right]{$s$} ;
    \draw (5,3.5) -- (4,3.5) node[below left]{$s-1$} ;
    \draw (4.5,4.5) node[above]{$+$} ;
    \draw (4.5,3.5) node[below]{$+$} ;
    \draw (4,4) node[left]{$-$} ;
    \draw (5,4) node[right]{$-$} ;
    
    \draw (4.5,2) node[above]{${b}(u;s)$} ;

    \draw (4,0) -- (4,1) node[above left]{$s$} ; 
    \draw (4,1) -- (5,1) node[above right]{$s-1$} ;
    \draw (5,1) -- (5,0) node[below right]{$s$} ;
    \draw (5,0) -- (4,0) node[below left]{$s+1$} ;
    \draw (4.5,1) node[above]{$-$} ;
    \draw (4.5,0) node[below]{$-$} ;
    \draw (4,0.5) node[left]{$+$} ;
    \draw (5,0.5) node[right]{$+$} ;  
    
    \draw (4.5,-1.5) node[above]{$\bar{b}(u;s)$} ;

    \draw (8,3.5) -- (8,4.5) node[above left]{$s$} ; 
    \draw (8,4.5) -- (9,4.5) node[above right]{$s-1$} ;
    \draw (9,4.5) -- (9,3.5) node[below right]{$s$} ;
    \draw (9,3.5) -- (8,3.5) node[below left]{$s-1$} ;
    \draw (8.5,4.5) node[above]{$-$} ;
    \draw (8.5,3.5) node[below]{$+$} ;
    \draw (8,4) node[left]{$-$} ;
    \draw (9,4) node[right]{$+$} ;
    
    \draw (8.5,2) node[above]{${c}(u;s)$} ;

    \draw (8,0) -- (8,1) node[above left]{$s$} ; 
    \draw (8,1) -- (9,1) node[above right]{$s+1$} ;
    \draw (9,1) -- (9,0) node[below right]{$s$} ;
    \draw (9,0) -- (8,0) node[below left]{$s+1$} ;
    \draw (8.5,1) node[above]{$+$} ;
    \draw (8.5,0) node[below]{$-$} ;
    \draw (8,0.5) node[left]{$+$} ;
    \draw (9,0.5) node[right]{$-$} ;  
    
    \draw (8.5,-1.5) node[above]{$\bar{c}(u;s)$} ;      
\end{tikzpicture}

The $R$-matrix \eqref{R-mat}  satisfies the dynamical Yang-Baxter equation \cite{GerN84,Fel95} (which is equivalent to Baxter's star-triangle relation for the Boltzmann weights $W$) :
\begin{multline}\label{YB}
  R_{12}(u_1-u_2; s+h_3) \; R_{13}(u_1-u_3 ; s) \; R_{23}( u_2-u_3 ; s+h_1) \\
  =
  R_{23}(u_2-u_3; s) \; R_{13}(u_1-u_3 ; s+h_2) \; R_{12}( u_1-u_2 ; s) ,
\end{multline}
with
\begin{equation}
   h=\begin{pmatrix} 1 & 0 \\ 0 & -1 \end{pmatrix}.
\end{equation}
This equation should be understood as an identity for meromorphic functions of $u_1,u_2,u_3,s$ with values in $\mathrm{End}(V\otimes V \otimes V)$.
The indices label as usual the space of the tensor product on which the corresponding operator acts. 
The $R$-matrix \eqref{R-mat} also satisfies the following properties:
\begin{itemize}\label{three-prop}
\item zero weight: $[ R_{12}(u;s) \, , \, h_1+h_2] =0 $,
\item unitarity: $R_{12}(u;s)\, R_{21}(-u;s)= \text{Id}$,
\item crossing symmetry: $\displaystyle \sigma_1^y R_{12}(-u - 1; s - h_1)  \sigma_1^y \, \frac{[s+h_2] [u]}{[s]\,[u+1]} =  R^{t_1}_{21}(u; s)$,\smallskip\\
in which, in the expression $R_{12}(-u - 1; s - h_1)$, the $h_1$ operator should be understood as acting to the right of any other operator involved in the definition of the $R$-matrix.
\end{itemize}

The transfer matrix of the model corresponds to the product of all statistical weights along a column of elementary cells of the lattice.
A given allowed configuration of heights along a vertical line of vertices of the lattice (a state on which the transfer matrix acts) corresponds to a $N$-tuple of heights $(s_1,s_2,\ldots,s_{N})$ such that $s_{i+1}-s_i=\pm 1$, $i=1,\ldots N$ (with the convention $s_{N+1}=s_1$), i.e. to a configuration $(s,\e_{\eps_1},\ldots,\e_{\eps_N})$ with $s=s_1$ and $\eps_i\equiv s_{i+1}-s_i$ such that $\eps_1+\dots+\eps_N=0$.
Hence, the space of states of the model can be seen as the space of functions $\mathrm{Fun}(\mathcal{H}[0])$ of one complex variable (the height $s$) with values in the zero-weight space $\mathcal{H}[0]$, where $\mathcal{H}=V^{\otimes N}$.

In the general (unrestricted) SOS model, $\eta$ is arbitrary and the dynamical parameter $s$ may take an infinite discrete set of values, i.e. belongs to some set $\mathbf{C_{s_0}}=s_0+\mathbb{Z}$ for some arbitrary parameter $s_0$.
Hence, in that case, the space of states is infinite-dimensional even for a finite lattice.
Usually, however, one considers situations for which 
the height $s$ is only allowed to take a finite set of values and the space of states is finite-dimensional.
This happens for rational values of $\eta$, i.e. for $\eta=r/L$ with $r$, $L$ being relatively prime integers such that $L>0$ (or more generally\footnote{This situation is actually equivalent to the case $\eta=r/L$ with $r$ being the greatest common divisor of $r_1$ and $r_2$ (see Appendix~\ref{ap-theta}).}, as in \cite{Bax73a},
when there exist two integers $r_1$ and $r_2$ such that $L\eta=r_1+r_2\tau$). In that case, that will be referred to as the {\em cyclic case} (or cyclic SOS (CSOS) model \cite{PeaS89}), the statistical weights of the model (i.e. the elements \eqref{bc1}-\eqref{bc2} of the $R$-matrix) are periodic of period $L$, and 
the space of states $\mathbf{H_{s_0}^L}$ corresponds to the space of functions  $\psi: \mathbf{C_{s_0}}\mapsto \mathcal{H}[0]$ such that $\psi(s+L)=\psi(s)$
or, equivalently,  to the space of functions  $\psi: \mathbf{C_{s_0}^L}\mapsto \mathcal{H}[0]$, with $\mathbf{C_{s_0}^{L}}=s_0+\mathbb{Z}/L\mathbb{Z}$.


The algebraic Bethe Ansatz approach to the SOS model, which enables one to diagonalize the transfer matrix of the model, has been developed in \cite{FelV96b}, based on the study of representations \cite{FelV96a} of Felder's dynamical quantum group $E_{\tau,\eta}(sl_2)$ associated to the dynamical $R$-matrix \eqref{R-mat}.
The central object of this approach is the monodromy matrix, defined as the following ordered product of $R$-matrices along a column of elementary cells of the lattice:
\begin{align}\label{monodromy}
 T_{a, 1\ldots N}(u ; \xi_1,\ldots, \xi_N ; s)
  &= R_{a N} (u - \xi_N ; s + h_1 +\cdots + h_{N-1} ) \ldots    R_{a 1} (u - \xi_1 ; s  ) \nonumber\\
  &= \begin{pmatrix} A(u ; s) & B(u ; s) \\
                                    C(u ; s) & D(u ; s) \end{pmatrix}_{\!\! [a]}.   
\end{align}
It acts on $V_a\otimes \mathcal{H}$, where $V_a\sim\mathbb{C}^2$ is another copy of $V$ usually called auxiliary space.
In this framework, the entries $A,B,C,D$ of the monodromy matrix are linear operators acting on $\mathcal{H}$, and the commutation relations of these operators are given by the following quadratic equation on $V_{a_1} \otimes V_{a_2} \otimes \mathcal{H}$,
\begin{multline}\label{RTT}
R_{a_1 a_2} (u_1-u_2 ; s + h_{1\ldots N}) \;
        T_{a_1, 1\ldots N}(u_1 ; s) \;  T_{a_2, 1\ldots N}(u_2 ; s+h_{a_1}) \\
= T_{a_2, 1\ldots N}(u_2 ; s)\;  T_{a_1, 1\ldots N}(u_1 ; s+h_{a_2}) \;
   R_{a_1 a_2} (u_1-u_2 ; s),
\end{multline}
which is a consequence of the Yang-Baxter relation \eqref{YB}.
Here $h_{1\ldots N} = h_1 + \ldots + h_N$.

It may be convenient to define, from the above monodromy matrix $T\in\mathrm{End}(V_a\otimes\mathcal{H})$, the following 
matrix of finite-difference operators:
\begin{equation}\label{mon-op}
   \widehat{T}(u)= \begin{pmatrix} \widehat{A}(u) & \widehat{B}(u) \\
                                    \widehat{C}(u) & \widehat{D}(u) \end{pmatrix}_{\!\! [a]} 
                            = T(u; \widehat{s}) \, \begin{pmatrix} \widehat{\tau}_s & 0 \\ 0 & \widehat{\tau}_s^{-1}\end{pmatrix}_{\!\! [a]}       
           \quad \in\mathrm{End}(V_a\otimes\mathrm{Fun}(\mathcal{H})) ,                
\end{equation}
where $\widehat{\tau}_s\, \widehat{s}=(\widehat{s}+1)\, \widehat{\tau}_s$, and the action of $\widehat{s}$ and $\widehat{\tau}_s$  on functions $f\in \mathrm{Fun}(\mathcal{H})$ are given as
\begin{equation}
    [\widehat{s} f](s)= s f(s), \qquad [\widehat{\tau}_s f](s)= f(s+1).
\end{equation}
This defines an operator algebra whose commutation relations follow from \eqref{RTT} (see \cite{FelV96a,FelV96b}). In this picture, the (operator) transfer matrix is  $\widehat{t}(u)=\widehat{A}(u)+\widehat{D}(u)$. It is easy to see \cite{FelV96a,FelV96b} that these transfer matrices preserve the space $\mathrm{Fun}(\mathcal{H}[0])$ of functions with values in the zero weight space $\mathcal{H}[0]$ (the subspace of $\mathcal{H}$ on which $h_{1\ldots N}$ vanishes), and that they commute pairwise on $\mathrm{Fun}(\mathcal{H}[0])$: $[\widehat{t}(u),\widehat{t}(v)]=0$ on $\mathrm{Fun}(\mathcal{H}[0])$.

The structure of the $R$-matrix implies that there exists a reference state $\ket{0} = \e_+\otimes\dots\otimes \e_+\in\mathcal{H}$ which is an eigenstate of $A(u;s)$ and $D(u;s)$. More precisely,
\begin{equation}
  A(u;s) \ket{0} = a(u) \ket{0}, \qquad D(u;s) \ket{0} = \frac{[s-1]}{[s+N-1]}\, d(u) \ket{0},
\end{equation}
with, in our normalization, 
\begin{equation}\label{a-d}
  a(u)=1,\qquad
   d(u) = \prod_{j=1}^N \frac{[u-\xi_j]}{[u-\xi_j+1]}.
\end{equation}
Starting from this reference state, one can construct states in the zero-weight space  $\mathrm{Fun}(\mathcal{H}[0])$ as
\begin{equation}\label{state}
 s \mapsto 
 \varphi(s) B(v_1;s) B(v_2;s-1)\ldots B(v_n;s-n+1) \ket{0},
\end{equation}
where $v_1,\ldots,v_n$ are arbitrary spectral parameters, $\varphi$ is an arbitrary numerical function of the dynamical parameter and $n$ is such that $N=2n+\aleph L$ for some integer $\aleph$ (or $N=2n$ in the unrestricted case). 
Bethe states correspond to states \eqref{state} for which the function $\varphi$ takes the form
 \begin{equation}\label{phi-FV}
   \varphi_\omega(s) = \omega^s \prod_{j=1}^{n} \frac{[1]}{[s-j]} , 
 \end{equation}
depending on some complex parameter $\omega$. Such states will be denoted as $\ket{\{v_1,\ldots,v_n\},\omega}$, or simply as $\ket{\{v\},\omega}$ when there is no ambiguity concerning the set $\{v\}\equiv\{v_1,\ldots ,v_n\}$.
Note that, whereas $\omega$ can a priori take any complex value in the unrestricted case, it is no longer the case in the cyclic case: for 
such a state $\ket{\{v\},\omega}$ to belong to $\mathbf{H_{s_0}^L}$ (i.e. to be $L$-periodic as a function of $s$), one should impose $\omega$ to be such that $(-1)^{rn} \omega^L=1$.

It can easily be shown, using the commutation relations coming from \eqref{RTT},
that, if $v_1,\ldots , v_n$ are off-diagonal\footnote{\label{foot-odiag}i.e. such that, for all $i<j$, $\eta v_i\neq\eta v_j\mod\mathbb{Z}+\tau\mathbb{Z}$.} solutions to the following system of Bethe equations\footnote{These equations are also valid in the unrestricted case for $N=2n$ with the convention $r\aleph=0$.}
 \begin{equation}\label{Bethe}
    a(v_j)  \prod_{l\ne j} \frac{[v_l-v_j+1]}{[v_l-v_j]} 
    =  (-1)^{r\aleph} \omega^{-2} \; d(v_j)  \prod_{l\ne j} \frac{[v_j-v_l+1]}{[v_j-v_l]} ,
    \quad j=1,\ldots n,
 \end{equation}
 %
then
\begin{equation}\label{act-transfer}
   \widehat{t}(u)\, \ket{\{ v\},\omega }= \tau (u; \{ v\},\omega )\, \ket{\{ v\},\omega },
\end{equation}
%
with
\begin{equation}
   \tau (u; \{v\},\omega)
    = \omega \; a(u) \prod_{l=1}^n \frac{[v_l-u+1]}{[v_l-u]}
        +  (-1)^{r\aleph}  \omega^{-1} \; d(u) \prod_{l=1}^n \frac{[u-v_l+1]}{[u-v_l]}.
    \label{tau}
\end{equation}

Similarly, setting $\bra{0}=\ket{0}^\dagger$ and
\begin{equation}\label{tphi-FV}
    \widetilde{\varphi}_\omega(s) =  {\omega}^{- s}  \prod_{j=0}^{n-1} \frac{[s+j]}{[1]},
\end{equation}
one can define Bethe states $\bra{\{v\},\omega}$ in the dual space of states as
\begin{equation}\label{dual}
   \bra{\{ v\} ,\omega}:
   s\mapsto \bra{0}C(v_n;s-n) \ldots  C(v_2;s-2) C(v_1;s-1)\widetilde{\varphi}_\omega(s).
\end{equation}
Then, if the set of spectral parameters $\{v\}\equiv\{v_1,\ldots , v_n\}$ is an off-diagonal solution of the system of Bethe equations \eqref{Bethe}, one has
\begin{equation}
 \bra{ \{v\}, \omega }\,\widehat{t} (u) = \tau(u; \{v\},\omega )\,  \bra{ \{v\} ,\omega } \label{act-transfer-dual},
\end{equation}
with $\tau(u; \{v\},\omega )$ given by \eqref{tau}.
%

\section{Scalar product of Bethe states}
\label{sec-sc-pdt}

The next step towards the calculation of form factors and correlation functions, after the determination of the eigenvectors of the transfer matrix, is to obtain manageable and compact formulas for their scalar products.
To compute these scalar products we use here the method proposed in \cite{KitMT99}, which is based on some recursion relation initiated from the determinant representation of the partition function with domain wall boundary conditions \cite{Ize87}. This recursion uses the expression of the operators $B$ and $C$ in the so-called $F$-basis, a basis of the space of states in which these operators become quasi-local \cite{MaiS00}. 

In the case of the dynamical SOS model, the $F$-basis has been explicitly obtained in \cite{AlbBFPR00}, whereas an expression (unfortunately not as a single determinant but as a sum of determinants) for the partition function of the model with domain wall boundary conditions has been obtained in \cite{PakRS08,Ros09} (see Appendix~\ref{app-part-fct}).
The fact that one does not know a compact formula (as a single determinant) for the partition function is a difficulty for the computation of the scalar product, but the approach of \cite{KitMT99} can nevertheless be achieved and, at least in the cyclic case for which the space of states is finite-dimensional, the result can be presented as a single determinant.

Let $\bra{\{u\},\omega_u}$ be a dual Bethe eigenstate given in terms of a function $\widetilde{\varphi}_{\omega_u}$ of the fom \eqref{tphi-FV} with complex parameter $\omega_u$, $\{u\}\equiv\{u_1,\ldots,u_n\}$ being solution of the corresponding Bethe equations,  and $\ket{\{v\},\omega_v}\in \mathbf{H_{s_0}^L}$ be a state of the form \eqref{state} given in terms of a function $\varphi_{\omega_v}$ \eqref{phi-FV} with complex parameter $\omega_v$, $\{v\}\equiv\{v_1,\ldots,v_n\}$ being a set of arbitrary parameters.
In the cyclic case, we consider the following scalar product between these two states:
\begin{align}
  \mathbf{S}_n(\{u\},\omega_u;\{v\},\omega_v) 
  &= \moy{ \{u\},\omega_u \mid \{v\},\omega_v } \nonumber\\
  &= \frac{1}{L}\sum_{s\,\in \, \mathbf{C_{s_0}^{L}} }
         \widetilde\varphi_{\omega_u}(s)\,\varphi_{\omega_v}(s) \, S_n ( \{u\}; \{v\} ; s)
  \nonumber\\
  &= \frac{1}{L} \sum_{s\,\in \, \mathbf{C_{s_0}^{L}} }
         \omega_u^{-s}\omega_v^s
        \prod_{j=1}^n\frac{[s+j-1]}{[s-j]} \, S_n ( \{u\}; \{v\} ; s),\label{scalarproduct}
\end{align}
where $S_n ( \{u\}; \{v\} ; s)$ is the partial scalar product at a given height $s$ defined as
\begin{equation}\label{Sn}
S_n ( \{u\}; \{v\} ; s)
  = \bra{0} C(u_{n};s-n)\ldots  C(u_{1};s-1)
                  B(v_{1};s) \ldots B(v_{n};s-n+1) \ket{0}.
\end{equation}
                  
Following the method proposed in \cite{KitMT99}, \eqref{Sn} can be calculated in the $F$-basis because the reference state and the dual reference state are left invariant under the action of the $F$-matrix which induces the corresponding change of basis (see \cite{MaiS00,KitMT99,AlbBFPR00} for more details about the $F$-basis). \eqref{Sn} is therefore equal to
\begin{equation}\label{Sntilde}
S_n ( \{u\}; \{v\} ; s)
  = \bra{0} \widetilde{C}(u_{n};s-n)\ldots  \widetilde{C}(u_{1};s-1)
                  \widetilde{B}(v_{1};s) \ldots \widetilde{B}(v_{n};s-n+1) \ket{0},
\end{equation}
where $\widetilde{C}$ and $\widetilde{B}$ stand for the expressions of the corresponding operators $C$ and $B$ in the $F$-basis (see \cite{AlbBFPR00} for details):
 \begin{align}
 \widetilde{B}(u;s)
   & = \frac{[s-1]}{[s+h_{1\ldots N}]}
       \sum_{i=1}^N \sigma_i^- 
       \frac{[1] [s+\sum_{l\ne i} h_l+u-\xi_i]}
               {[s+\sum_{l\ne i} h_l ] [u-\xi_i+1]}
          \underset{j \ne i}{\otimes}
          \begin{pmatrix}
              \frac{[u-\xi_j]}{[u-\xi_j+1]} & 0 \\
              0 & \frac{[\xi_j-\xi_i+1]}{[\xi_j-\xi_i]}
          \end{pmatrix} _{\!\! [j]},
          \label{Btilde}\\
 \widetilde{C}(u;s)
 & = 
       \sum_{i=1}^N \sigma_i^+
       \frac{[1] [s-u+\xi_i]}
               {[s ] [u-\xi_i+1]}
          \underset{j \ne i}{\otimes}
          \begin{pmatrix}
              \frac{[u-\xi_j]}{[u-\xi_j+1]} \frac{[\xi_i-\xi_j+1]}{[\xi_i-\xi_j]} & 0 \\
              0 & 1
          \end{pmatrix} _{\!\![j]}.
          \label{Ctilde}
\end{align}
 
As already mentioned, \eqref{Sntilde} can be computed by recursion.
For $k\in\{0,1,\ldots n\}$, we consider the following quantity:
\begin{multline}\label{Gfunction}
 G^{(k)}_{\ell_{k+1},\ldots, \ell_n} ( \{u\}; \{v_{1},\ldots, v_{k} \}; s) \\
   = \bra{0} \widetilde{C} (u_{n};s-n) \ldots \widetilde{C}(u_{1};s-1)
                            \widetilde{B}(v_{1};s) \ldots  \widetilde{B}(v_{k};s-k+1)
                                    \ket{\ell_{k+1},\ldots , \ell_n},                                 
\end{multline}
where $\ket{\ell_{k+1},\ldots , \ell_n}$ denotes the state with $N-k$ down spins at sites $\ell_{k+1},\ldots , \ell_n$.
For $k=n$, the quantity $G^{(n)}( \{u\}; \{ v_{1},\ldots, v_{n} \} ; s)$ corresponds to the partial scalar product \eqref{Sntilde}.
For $k=0$, $G^{(0)}_{\ell_{1},\ldots, \ell_n} ( \{u\}; \emptyset ; s)$ can easily be computed, by means of the expression \eqref{Ctilde} of the operator $C$ in the $F$-basis, in terms of the partition function with domain wall boundary conditions \eqref{part-fct} on a lattice of size $n\times n$ (we also use the fact that the $F$-matrix leaves invariant the state $\ket{\bar{0} }$ so that \eqref{part-fct} can be written directly in the $F$-basis):
 \begin{align}
 G^{(0)}_{\ell_1,\ldots, \ell_n} ( \{u\}; \emptyset ;s)
  &= \bra{0} \widetilde{C}_{1\ldots N}(u_{n};s-n) \ldots
                                  \widetilde{C}_{1\ldots N}(u_{1};s-1) \ket{\ell_1,\ldots , \ell_n}
                                  \nonumber\\
  &=       \prod_{j\ne \ell_1,\ldots \ell_n} \left\{
                              \prod_{\alpha=1}^n \frac{ [u_{\alpha}-\xi_j] }{ [u_{\alpha}-\xi_j+1]} \cdot
                              \prod_{k=1}^n \frac{[\xi_{\ell_k}-\xi_j+1]}{[\xi_{\ell_k}-\xi_j]} \right\}
              \nonumber\\
  &\hspace{4cm}\times            
              Z_{n}(  u_1,\ldots, u_n ; \xi_{\ell_1},\ldots, \xi_{\ell_n} ; s-n).
              \label{initial}
\end{align}
The functions \eqref{Gfunction} admit the following recursion relation:
\begin{multline}
G^{(k)}_{\ell_{k+1},\ldots, \ell_n} ( \{u\}; \{v_{1},\ldots, v_{k} \} ; s) 
      = \sum_{\ell_k \ne \ell_{k+1}, \ldots, \ell_n}  \hspace{-2mm}     
      G^{(k-1)}_{\ell_{k},\ldots, \ell_n} ( \{u\}; \{v_{1},\ldots, v_{k-1} \}; s) \\
      \times
       \bra{\ell_k,\ldots,\ell_n}  \widetilde{B}(v_{k};s-k+1)
                                    \ket{\ell_{k+1},\ldots , \ell_n},
     \label{induction}                                
\end{multline}
with, using the expression \eqref{Btilde} of the operator $B$ in the $F$-basis,
\begin{multline}
\bra{\ell_k,\ldots,\ell_n}  \widetilde{B}(v_{k};s-k+1)  \ket{\ell_{k+1},\ldots , \ell_n} 
    =   
          \frac{[1]\,[s+N-2n+k+v_{k}-\xi_{\ell_k}]}{[s+N-2n+k]\, [v_{k}-\xi_{\ell_k}]}  
         \\
         \times \frac{[s-k]\, d(v_k)}{[s+N-2n+k-1] }
         \prod_{j=k+1}^n \!\! \left\{ \frac{[v_{k}-\xi_{\ell_j}+1]}{[v_{k}-\xi_{\ell_j}]}
                                       \frac{[\xi_{\ell_j}-\xi_{\ell_k}+1]}{[\xi_{\ell_j}-\xi_{\ell_k}]} \right\}. 
         \label{coeff-rec}                                                                                                                          
 \end{multline}

The recursion relation \eqref{induction} with initial condition \eqref{initial} is explicitly solved in Appendix~\ref{app-sc}. Here we merely give the result for the partial scalar product \eqref{Sn}.
\begin{prop}\label{prop-sc}
For  $N=2n+\aleph L$ with $\aleph$ integer, let $\{u_{1}, \ldots , u_{n}\}$ be a solution of the Bethe equations \eqref{Bethe} with complex parameter $\omega_u$, and $\{v_{1},\ldots,v_{n}\}$ be a set of arbitrary parameters. Then the partial scalar product $S_n ( \{u\}; \{v\} ; s)$ can be represented as the following sum of determinants:
\begin{multline}\label{result-sp}
   S_n ( \{u\}; \{v\} ; s)
  = \frac{[s-n]}{ [\gamma]^n[ |u| -|v|+\gamma+s] }
      \prod_{j=1}^{n-1}\frac{[s-j]}{[s+j]}   \,   
     \frac{ \prod_{t=1}^{n}  d(u_{t})  \, d(v_{t}) }{ \prod_{j<k} [u_{j}-u_{k}] [v_{k}-v_{j}] }     
  \\
 \times \sum_{ S,\tilde{S}\subset\{1,\ldots, n\}  } \hspace{-2mm} (-1)^{|S|+|\tilde{S}|} \,
       \prod_{j\not\in\tilde{S}} \Bigg\{(-1)^{r\aleph}\,
  \frac{a(v_{j})}{d(v_{j}) } \prod_{t=1}^n [u_{t}-v_{j}+1] \Bigg\}
   \prod_{j\in\tilde{S}} \Bigg\{\omega_u^{-2} \prod_{t=1}^n [u_{t}-v_{j}-1] \Bigg\}
   \\
      \times  \frac{[\gamma+s-|S|+|\tilde{S}| ]}{[s-|S|+|\tilde{S}| ]}\,
         \det_n \big[ \mathcal{N}_\gamma (\{u\}; \{ v-\delta^{S\,\tilde{S} } \}  ) \big]  .
\end{multline}
In this expression, $\gamma$ is an arbitrary complex parameter (the result does not depend on $\gamma$), $|u|=u_1+\dots+u_n$, $|v|=v_1+\dots+v_n$, and the elements of the $n\times n$ matrix $\mathcal{N}_\gamma$ are given by
\begin{equation}\label{det-N}
   \big[ \mathcal{N}_\gamma (\{u\}; \{v \}  ) \big]_{j k} =
    \frac{[u_{j}-v_{k}+\gamma]}{[u_{j}-v_{k}]}.
\end{equation}
The sum in \eqref{result-sp} runs over all subsets $S$ and $\tilde{S}$ of $\{1,\ldots,n\}$, $|S|$ and $|\tilde{S}|$ being the cardinality of these subsets, and $\{ v-\delta^{S\,\tilde{S} } \}=\{v_1-\delta^{S\,\tilde{S} }_1,\ldots, v_n-\delta^{S\,\tilde{S} }_n \}$, with
\begin{equation}
 \delta_{k}^{S \tilde{S}}=
  \begin{cases}
       1 & \text{if }\ k\in S\ \text{ and }\ k\notin\tilde{S}, \\
       -1 & \text{if }\ k\notin S\ \text{ and }\ k\in\tilde{S}, \\
       0      & \text{if }\ k\notin S\cup\tilde{S}\ \text{ or }\ k\in S\cap\tilde{S}.
  \end{cases}
 \end{equation}
\end{prop}

\begin{rem}
The formula \eqref{result-sp} is also valid for any generic SOS model (i.e. for generic $\eta$) in the case $N=2n$ ($\aleph=0$).
\end{rem}
\begin{rem}
The apparent poles at $v_k=u_j$ are removable due to the fact that $\{u\}$ satisfies the Bethe equations.
The apparent pole at  $|u| -|v|+\gamma+s=0$ is also removable due to Remark~\ref{rem-pole-part}.
\end{rem}

In the case when $\eta$ is rational, i.e. if there exists a positive integer $L$ such that $L\eta$ is integer, the sum of determinants in \eqref{result-sp} can be reduced, by a similar argument as for the partition function in \cite{Ros09}, to a sum over only $L$ terms. Indeed, using \eqref{sum-L} and the periodicity of the theta-function to re-express the ratio $\frac{[\gamma+s-|S|+|\tilde{S}| ]}{[s-|S|+|\tilde{S}| ]}$ in \eqref{result-sp}, one obtains the following result:
\begin{cor}
Suppose that there exists a positive integer $L$ such that $L\eta$ is integer. Then, with the same hypothesis and notations as in Proposition~\ref{prop-sc}, the partial scalar product $S_n(\{u\};\{v\};s)$ can be written as a sum of only $L$ terms:
\begin{multline}\label{result-spL}
   S_n ( \{u\}; \{v\} ; s)
  = \frac{[\gamma]\, [s]}{ [0]'\,[ |u| -|v|+\gamma+s] }
      \prod_{j=1}^{n}\frac{[s-j]}{[s+j-1]}   \,   
     \frac{ \prod_{t=1}^{n}  d(u_{t})   }{ \prod_{j<k} [u_{j}-u_{k}] [v_{k}-v_{j}] }     
  \\
 \times \sum_{ \ell=0  }^{L-1} q^{\ell s}
         \frac{[Ls_0+\gamma+\ell\frac\tau\eta]_{_L}\, [0]'_{_L}}{[Ls_0]_{_L}\,[\gamma+\ell\frac\tau\eta]_{_L}} \,
         \det_n \big[ \Omega_\gamma^{(\ell)} (\{u\},\omega_u; \{ v\}  ) \big]  ,
\end{multline}
with
\begin{multline}\label{def-OmegaL}
   [ \Omega_\gamma^{(\ell)} (\{u \},\omega_u; \{ v \}  ) \big]_{ij} 
   = \frac{(-1)^{r\aleph} }{[\gamma]}\Bigg\{\! \frac{[u_{i}-v_{j}+\gamma]}{[u_{i}-v_{j}]}
                  -q^{-\ell}\frac{[u_{i}-v_{j}+\gamma+1]}{[u_{i}-v_{j}+1]} \! \Bigg\} \,
       a(v_{j}) \! \prod_{t=1}^{n} [u_{t}-v_{j}+1]
       \\
     + \frac{1}{[\gamma]}
     \Bigg\{\! \frac{[u_{i}-v_{j}+\gamma]}{[u_{i}-v_{j}]}
                  -q^\ell\frac{[u_{i} -v_{j}+\gamma-1]}{[u_{i}-v_{j}-1]} \!  \Bigg\}    \,
        \omega_u^{-2} d(v_{j}) \! \prod_{t=1}^{n} [u_{t}-v_{j}-1]. 
\end{multline}
Here we have set $q=e^{2\pi i\eta}$ and $[u]_{_L}=\theta_1(\eta u; L\tau)$.
\end{cor}
Although we do not obtain a single determinant, the representation \eqref{result-spL} seems nevertheless more convenient than \eqref{result-sp} for the computation of the correlation functions, its main advantage being that the number of terms remains finite in the thermodynamic limit.

\bigskip

At this point, we would like to make the following remark. As we have seen, we did not succeed to represent the partial scalar product \eqref{Sn} as a single determinant as in \cite{Sla89}. However, it is worth stressing that the state
\begin{equation}\label{partial-state}
  B(v_1;s) B(v_2;s-1)\ldots B(v_n;s-n+1)\ket{0} \in \mathcal{H}[0]
\end{equation}
is {\em not} a Bethe state: even when the parameters $\{v\}$ satisfy the Bethe equations, the shifts of the dynamical parameter that appear when commuting $A(u;s)$ or $D(u;s)$ with $B(u;s\pm 1)$ prevent this state from being an eigenstate of, for instance, $A(u;s)+D(u;s)$. Therefore, it is not very surprising that the occurrence of dynamical shifts also prevent us from re-expressing the quantity \eqref{result-sp} or \eqref{result-spL} (and the partition function \eqref{part-fct} as a particular case) as a single determinant as in the non-dynamical case.
In fact, if we consider instead the true Bethe state, i.e. the {\em function} \eqref{state}, we shall see that the natural scalar product of functions \eqref{scalarproduct} can, in the cyclic case, and for one of the set of parameters being solution of the Bethe equations, be represented as a single determinant.

\bigskip
 
Let us therefore now consider the scalar product \eqref{scalarproduct}.
Using the $L$-periodicity and the fact that the right hand side of \eqref{result-sp} does not depend on $\gamma$ and hence that a convenient choice can be made for this parameter, we can extract the dependance in the dynamical parameter $s$ from the sum of determinants and factorize the latter as a single determinant. More precisely, we get from \eqref{result-sp}
\begin{align}
  \mathbf{S}_n(\{u\},\omega_u;\{v\},\omega_v) 
    &= \frac{1}{ [\gamma]^n }    
     \frac{ \prod_{t=1}^{n}  d(u_{t})  \, d(v_{t}) }{ \prod_{j<k} [u_{j}-u_{k}] [v_{k}-v_{j}] }
          \sum_{ S,\tilde{S}\subset\{1,\ldots, n\}  }  (-1)^{|S|+|\tilde{S}|}       
  \nonumber\\
 &\times
 \prod_{j\not\in\tilde{S}} \Bigg\{(-1)^{r\aleph}
  \frac{a(v_{j})}{d(v_{j}) } \prod_{t=1}^n [u_{t}-v_{j}+1] \Bigg\}
   \prod_{j\in\tilde{S}} \Bigg\{\omega_u^{-2} \prod_{t=1}^n [u_{t}-v_{j}-1] \Bigg\}
   \nonumber\\
      &\times
 \Bigg\{  \frac{1}{L} \sum_{s\,\in \, \mathbf{C_{s_0}^{L}} }
 \frac{\omega_u^{-s}\omega_v^s \,[ s]}{ [ |u| -|v|+\gamma+s] }
   \frac{[\gamma+s-|S|+|\tilde{S}| ]}{[s-|S|+|\tilde{S}| ]} \Bigg\}
   \nonumber\\
     & \hspace{5cm} \times
         \det_n \big[ \mathcal{N}_\gamma (\{u\}; \{ v-\delta^{S\,\tilde{S} } \}  ) \big]  .      
\end{align}
Setting $\gamma=-|u| +|v|$, and performing a change of indices in the sum over $s$, we obtain
\begin{align}
  \mathbf{S}_n(\{u\},\omega_u;\{v\},\omega_v) 
     &= 
     \Bigg\{  \frac{1}{L} \sum_{s\,\in \, \mathbf{C_{s_0}^{L}} }
    \omega_u^{-s}\omega_v^{s}
   \frac{[\gamma+s]}{[s]} \Bigg\}\,
\frac{1}{ [\gamma]^n }    
     \frac{ \prod_{t=1}^{n}  d(u_{t})  \, d(v_{t}) }{ \prod_{j<k} [u_{j}-u_{k}] [v_{k}-v_{j}] }
  \nonumber\\
 &\hspace{-3mm}\times
    \sum_{ S,\tilde{S}\subset\{1,\ldots, n\}  } \hspace{-3mm} (-1)^{|S|+|\tilde{S}|} \,
    \Big(  \frac{\omega_v}{\omega_u}\Big)^{|S|- |\tilde{S}|} 
        \prod_{j\not\in\tilde{S}} \Bigg\{(-1)^{r\aleph}
  \frac{a(v_{j})}{d(v_{j}) } \prod_{t=1}^n [u_{t}-v_{j}+1] \Bigg\}
   \nonumber\\
     &\hspace{-3mm}  \times   
   \prod_{j\in\tilde{S}} \Bigg\{ \omega_u^{-2} \prod_{t=1}^n [u_{t}-v_{j}-1] \Bigg\}
     \det_n \big[ \mathcal{N}_\gamma (\{u\}; \{ v-\delta^{S\,\tilde{S} } \}  ) \big] .
        \label{S-intermediate}
\end{align}
Using the linearity of the determinant, one can now express the last sum in \eqref{S-intermediate} as a single determinant.
We have the following result:
\begin{thm}    
For  $N=2n+\aleph L$ with $\aleph$ integer, let $\{u_{1}, \ldots , u_{n}\}$ be a solution of the Bethe equations \eqref{Bethe} with complex parameter $\omega_u$, and $\{v_{1},\ldots,v_{n}\}$ be a set of arbitrary parameters. Then, the scalar product $\mathbf{S}_n ( \{u\},\omega_u; \{v\},\omega_v)$ \eqref{scalarproduct} between the Bethe eigenstate $\bra{\{u\},\omega_u }$ and the state $\ket{\{v\},\omega_v }\in \mathbf{H_{s_0}^L}$ of the form \eqref{state},\eqref{phi-FV} with parameter $\omega_v$ can be represented as:
\begin{multline}\label{scal-prod}
  \mathbf{S}_n ( \{u\},\omega_u; \{v\},\omega_v) 
         = 
     \Bigg\{   \frac{1}{L} \sum_{s\,\in \, \mathbf{C_{s_0}^{L}} }
                    \frac{\omega_v^s}{\omega_u^s} \frac{[\gamma+s]}{[s]} \Bigg\}\,
     \frac{ \prod_{t=1}^{n}    d(u_{t})  
               }
            { \prod_{j<k} [u_{j}-u_{k}] [v_{k}-v_{j}] }
            \\
            \times\det_n \big[ \Omega_{\gamma} (\{u\},\omega_u; \{ v \} ,\omega_v ) \big],                        
\end{multline}
with $\gamma=-|u| +|v|$ and $\Omega_\gamma$ given by
\begin{multline}\label{def-Omega}
   \hspace{-2mm}[ \Omega_\gamma (\{u \},\omega_u; \{ v \},\omega_v  ) \big]_{ij} 
   = \frac{(-1)^{r\aleph} }{[\gamma]}\Bigg\{\! \frac{[u_{i}-v_{j}+\gamma]}{[u_{i}-v_{j}]}
                  -\frac{\omega_v}{\omega_u}\frac{[u_{i}-v_{j}+\gamma+1]}{[u_{i}-v_{j}+1]} \! \Bigg\} 
       a(v_{j}) \! \prod_{t=1}^{n} [u_{t}-v_{j}+1]
       \\
     + \frac{1}{[\gamma]}
     \Bigg\{\! \frac{[u_{i}-v_{j}+\gamma]}{[u_{i}-v_{j}]}
                  -\frac{\omega_u}{\omega_v}\frac{[u_{i} -v_{j}+\gamma-1]}{[u_{i}-v_{j}-1]} \!  \Bigg\}    
        \omega_u^{-2} d(v_{j}) \! \prod_{t=1}^{n} [u_{t}-v_{j}-1]. 
\end{multline}
\end{thm} 

From this determinant formula, it can be shown (see Appendix~\ref{app-orth}) that two different Bethe eigenstates are orthogonal.
On the contrary, taking the limit in which $\{v\}=\{u\}$ with $\omega_v=\omega_u$  in \eqref{scal-prod}, we obtain 
the formula for the ``square of the norm'' of a Bethe eigenstate
(in that case $\gamma=0$):
\begin{align}
  \mathbf{S}_n(\{u\},\omega_u;\{u\},\omega_u) 
     &= 
     \frac{ \prod_{t=1}^{n}    d(u_{t}) \cdot 
               \det_n \big[ \Omega_{0} (\{u\},\omega_u; \{ u \},\omega_u  ) \big] }
            { \prod_{j \neq k} [u_{j}-u_{k}]  }
            \nonumber\\
     &=        
     \frac{ (-1)^{nr\aleph} }{(-[0]')^n}
     \frac{ \prod_{t=1}^{n}  a(u_{t})  d(u_{t}) \,
                    \prod_{j,k=1}^n [u_{j}-u_{k}+1] }
             { \prod_{j\not= k} [u_{j}-u_{k}]  }   
     \det_n \big[ \Phi (\{u\} ) \big]   ,\label{gaudin}
\end{align}  
with
\begin{equation}
   \big[ \Phi (\{u\} ) \big]_{jk}
   = \delta_{jk}\Bigg\{ \log'\frac{a}{d}(u_{j})
                                    +\sum_{t=1}^n \tilde{K}(u_{j}-u_{t})\Bigg\}
      - \tilde{K}(u_{j}-u_{k}).
\end{equation}   
Here we have defined the even function
\begin{align}
   \tilde{K}(u)= \frac{[u-1]'}{[u-1]}-\frac{[u+1]'}{[u+1]}.
\end{align}
Note that the formula \eqref{gaudin} is very similar to its XXZ analog \cite{GauMcCW81,Kor82}.

\begin{rem}
In the case $\aleph=0$ ($n=N/2$), the above formula does not explicitly depend on the periodicity $L$ of the model. Hence, one can easily take the limit $L\to\infty$, extending by continuity the validity of \eqref{gaudin} to all values of $\eta$, including irrational ones.
In that case, the cyclic condition has only been used so as to avoid the subtleties of dealing with an infinite-dimensional space of states.
\end{rem}

\section{Solution of the quantum inverse problem}
\label{sec-inv}

In the ABA framework, the Bethe states are constructed by a repeated action of $n$ non-local operators $B$ on the reference state $\ket{0}$. In order to compute form factors and correlation functions, one needs to be able to act explicitly  with local operators on such states, which may seem not so easy (at least by direct computation) due to the mixture of local and non-local operators in  a same expression. A way to solve this problem was proposed in \cite{KitMT99}: it relies on the solution of the quantum inverse problem, which consists in expressing the local operator we consider in terms of the generators of the Yang-Baxter algebra; as a result, it is possible to compute the action of this local operator on a Bethe state by using only the quadratic Yang-Baxter commutation relations given by the $R$-matrix.

The method described in \cite{KitMT99,MaiT00}, which relies on the fundamental property that the $R$-matrix evaluated at $0$ coincides with a permutation operator, can also be applied  to the dynamical case. Indeed, the structure of the $R$-matrix \eqref{R-mat} implies that, for any value of $s$,
\begin{equation}\label{R-perm}
 R_{ij}(0;s) = P_{ij},
\end{equation}
where $P_{ij}$ is the permutation operator of spaces $i$ and $j$.
The explicit reconstruction of local operators in the dynamical SOS case involves however some subtleties (with respect to the simplest XXZ case) due to the presence of the dynamical parameter $s$.

Let $E_i^{\alpha \beta}$ be the elementary matrix, acting on the  $i$-th space of the tensor product $\mathcal{H}=V^{\otimes N}$, with elements $(E_i^{\alpha \beta})_{jk}=\delta^\alpha_j\delta^\beta_k$, where $\alpha$ and $\beta$ are equal to $\pm 1$.
In order to adapt the method of \cite{KitMT99,MaiT00} to the dynamical case, we first note that
\begin{equation}\label{E-h}
    h_i E_i^{\alpha \beta} = E_i^{\alpha \beta} (h_i + \alpha-\beta).
\end{equation}
We shall also use the following lemmas:
\begin{lemma}\label{lem-id-T}
For any value of $s$, we have the identity
\begin{equation}\label{id-T}
  T_{a_1,1\ldots N}(\xi_1;s) \, T_{a,1\ldots N}(u;s+h_{a_1})
  = T_{a,2\ldots N a_1}(u;s+h_1) \, T_{a_1,1\ldots N}(\xi_1;s+h_a),
 \end{equation} 
in which $T_{a,1\ldots N}$ (respectively $T_{a,2\ldots N a_1}$) is the monodromy matrix of a chain of $N$ sites labelled (in this order) by $1,2,\ldots N$ (respectively by $2,3,\ldots N,a_1$) with inhomogeneity parameters $\xi_1,\xi_2,\ldots \xi_N$ (respectively $\xi_2,\xi_3,\ldots \xi_N,\xi_1$). 
\end{lemma}
\Proof
Applying the quadratic commutation relation \eqref{RTT} on the l.h.s. of \eqref{id-T}, we get
\begin{multline}
  T_{a_1,1\ldots N}(\xi_1;s) \, T_{a,1\ldots N}(u;s+h_{a_1})
  = R_{a_1 a}^{-1}(\xi_1-u;s+ h_{1\ldots N})\\
  \times T_{a,1\ldots N}(u;s) \, T_{a_1,1\ldots N}(\xi_1;s+h_a) \,
    R_{a_1 a} (\xi_1-u;s).
\end{multline}
Then, using the fact that $R_{a_1 1}(0)= P_{a_1 1}$ and that $R$ is of weight 0, we have
\begin{equation}
 T_{a_1,1\ldots N}(\xi_1;s+h_a) \, R_{a_1 a} (\xi_1-u;s)
 = R_{1 a} (\xi_1-u;s) \, T_{a_1,1\ldots N}(\xi_1;s+h_a).
\end{equation}
On the other hand, the $R$-matrix being unitary, we have
\begin{equation}
  R_{a_1 a}^{-1}(\xi_1-u;s+ h_{1\ldots N}) \, T_{a,1\ldots N}(u;s) \, R_{1 a} (\xi_1-u;s)
  = T_{a, 2\ldots N a_1} (u; s+h_1),
\end{equation}
which ends the proof.
\qed

\begin{lemma}\label{lem-id-T2}
For any value of $s$, we have the following identity between products of monodromy matrices:
\begin{multline}\label{id-T2}
  T_{a_1,1\ldots N}(\xi_1;s) \, T_{a_2,1\ldots N}(\xi_2;s+h_{a_1}) \ldots
  T_{a_i,1\ldots N}(\xi_i;s+h_{a_1}+h_{a_2}+\cdots+h_{a_{i-1}}) \\
  = T_{a_i,i\ldots N a_1 a_2\ldots a_{i-1}}(\xi_i;s+h_1+h_2+\cdots +h_{i-1})\\
  \times
   T_{a_{i-1},i-1\ldots N a_1 a_2\ldots a_{i-2}}(\xi_{i-1};s+h_1+h_2+\cdots +h_{i-2}+h_{a_i})
  \ldots\\
  \ldots T_{a_2,2\ldots N a_1}(\xi_2;s+h_1+h_{a_3}+\cdots+h_{a_i}) \,
   T_{a_1,1\ldots N}(\xi_1;s+h_{a_2}+\cdots+h_{a_i}),
\end{multline} 
where the notations are similar to Lemma~\ref{lem-id-T} (i.e. for instance $T_{a_i,i\ldots N a_1 a_2\ldots a_{i-1}}$ denotes the monodromy matrix of a chain of $N$ sites labelled by $i,\ldots N, a_1, a_2,\ldots a_{i-1}$ with inhomogeneity parameters $\xi_{i},\xi_{i+1},\ldots\xi_N,\xi_1,\xi_2,\ldots\xi_{i-1}$).
\end{lemma}
\Proof
\eqref{id-T2} can be proven by induction on $i$ using \eqref{id-T}.
\qed

The solution of the quantum inverse problem, i.e. the reconstruction of local operators $E_i^{\alpha\beta}$ in terms of the entries of the monodromy matrix \eqref{mon-op}, can be formulated as follows:
\begin{thm}
  A local operator $E_i^{\alpha\beta}$ acting on the  $i$-th space of the tensor product $\mathcal{H}=V^{\otimes N}$ can be expressed in terms of the entries of the monodromy matrix \eqref{mon-op} in the following way:
  \begin{equation}\label{inv-pb1}
    E_i^{\alpha\beta}    
    = \prod_{k=1}^{i-1} \widehat{t}(\xi_k) \cdot \widehat{T}_{\beta \alpha} (\xi_i)
        \cdot \prod_{k=i}^1\big[ \widehat{t}(\xi_k) \big]^{-1} \cdot\, \widehat{\tau}_s^{\beta-\alpha}.
  \end{equation}
  Hence, a product of local operators on adjacent sites admits the following reconstruction:
  \begin{equation}\label{inv-pb2}
    E_i^{\alpha_i \beta_i}     E_{i+1}^{\alpha_{i+1}\beta_{i+1}}  \!  \ldots E_{i+j}^{\alpha_{i+j}\beta_{i+j}}  
    =   \prod_{k=1}^{i-1} \widehat{t}(\xi_k) \cdot
         \prod_{k=i}^{i+j}  \widehat{T}_{\beta_k \alpha_k} (\xi_k) \cdot
         \!\!\prod_{k=i+j}^1\!\!\big[ \widehat{t}(\xi_k) \big]^{-1} \!\cdot\, 
         \widehat{\tau}_s^{\sum_{k=i}^{i+j}[\beta_k-\alpha_k]}.
  \end{equation}
\end{thm}

\Proof
 Let us first prove \eqref{inv-pb1} for $i=1$.
 Writing $\widehat{T}_{\beta \alpha} (\xi_1)$ as a trace over an auxiliary space, and expressing the monodromy matrix \eqref{mon-op} in terms of products of $R$-matrices and shift operator $\widehat{\tau}_s$, we get
 \begin{align}
 \widehat{T}_{\beta \alpha} (\xi_1)
 &= \tr_a [\widehat{T}_{a,1\ldots N}(\xi_1;\xi_1,\ldots,\xi_N) \, E_a^{\alpha\beta}] \nonumber\\
 &=\tr_a [ R_{a N} (\xi_1 - \xi_N ; \widehat{s} + h_1 +\cdots + h_{N-1} ) \ldots    
                 R_{a 2} (\xi_1 - \xi_2 ; \widehat{s}+h_1  ) \,
                 P_{a1} \, \widehat{\tau}_s^{h_a}\, E_a^{\alpha\beta} ] , \nonumber
\end{align}
in which we have used \eqref{R-perm}.
Passing in this expression the operator $E_a^{\alpha\beta}$ from right to left  by means of  \eqref{E-h}, we obtain
\begin{align}
 \widehat{T}_{\beta \alpha} (\xi_1)
 &= \tr_a [ E_1^{\alpha\beta}\,     R_{a N} (\xi_1 - \xi_N ;  \widehat{s} + h_1 +\cdots + h_{N-1} +\alpha-\beta ) \ldots \nonumber\\ 
 &\hspace{5cm} \ldots    R_{a 2} (\xi_1 - \xi_2 ;  \widehat{s}+h_1 +\alpha-\beta  ) \, P_{a1} \,  \widehat{\tau}_s^{h_a+\alpha-\beta} ] 
                 \nonumber \\  
&=    E_1^{\alpha\beta}\, \tr_a[  T_{a,1\ldots N}(\xi_1;\xi_1,\ldots,\xi_N;  \widehat{s}+\alpha-\beta)\,  \widehat{\tau}_s^{h_a+\alpha-\beta} ]   \nonumber \\  
&=  E_1^{\alpha\beta}\, \widehat{\tau}_s^{\alpha-\beta}\, \widehat{t}(\xi_1),  
\label{reconstr1}                   
\end{align}
which ends the proof of \eqref{inv-pb1} in the case $i=1$. 

In the general case, one proceeds similarly by  writing $\prod_{k=1}^{i-1} \widehat{t}(\xi_k) \cdot \widehat{T}_{\beta \alpha} (\xi_i)$ as a trace over auxiliary spaces:
\begin{align} 
 \prod_{k=1}^{i-1} \widehat{t}(\xi_k) \cdot \widehat{T}_{\beta \alpha} (\xi_i)
  &=  \tr_{a_1 \ldots a_{i-1} a_i} [ \widehat{T}_{a_1}(\xi_1)\ldots \widehat{T}_{a_{i-1}}(\xi_{i-1})\, 
                                                              \widehat{T}_{a_{i}}(\xi_i)\, E_{a_i}^{\alpha\beta}]
                                                              \nonumber\\
   &=   \tr_{a_1 \ldots a_{i-1} a_i} [ {T}_{a_1}(\xi_1;\widehat{s})\ldots 
                                                            {T}_{a_{i-1}}(\xi_{i-1};\widehat{s}+h_{a_1}+\dots+h_{a_{i-2}})
                                                                \nonumber\\
   &\hspace{1.5cm}\times                                                             
                              {T}_{a_i}(\xi_i ;\widehat{s}+h_{a_1}+\dots+h_{a_{i-1}})\, 
                               \widehat{\tau}_s^{h_{a_1}+\dots + h_{a_{i-1}}+ h_{a_i} }\,E_{a_i}^{\alpha\beta}] .   
                               \label{reconstruct2}
\end{align}
Passing the elementary matrix $E_{a_i}^{\alpha\beta}$ from the right to the left, first through the shift operator $\widehat{\tau}_s^{h_{a_1}+\dots + h_{a_{i-1}}+ h_{a_i} }$, and then through the product of monodromy matrices, one obtains
\begin{align}
 &\prod_{k=1}^{i-1} \widehat{t}(\xi_k) \cdot \widehat{T}_{\beta \alpha} (\xi_i)
  = E_i^{\alpha\beta}    \,
         \tr_{a_1 \ldots a_{i-1} a_i} [  {T}_{a_1}(\xi_1;\widehat{s}+\alpha-\beta)\ldots 
        \nonumber \\
  &\hspace{2.8cm} \ldots    {T}_{a_{i-1}}(\xi_{i-1};\widehat{s}+h_{a_1}+\dots+h_{a_{i-2}}+\alpha-\beta)\,                                         
        \nonumber \\
  &\hspace{2.8cm} \times  {T}_{a_i}(\xi_i;\widehat{s}+h_{a_1}+\dots+h_{a_{i-1}}+\alpha-\beta)\, 
                                    \widehat{\tau}_s^{h_{a_1}+\dots + h_{a_{i-1}}+ h_{a_i} +\alpha-\beta}     ] .
                                    \label{reconstruct3}
\end{align}
Here we have used \eqref{E-h}, and the fact that the product of monodromy matrices could be expressed in a convenient way by means of Lemma~\ref{lem-id-T2}, so that its commutation with $E_a^{\alpha\beta}$ (which becomes $E_i^{\alpha\beta}$ by this process) can be performed similarly as in the case $i=1$.
Finally, it remains to note that \eqref{reconstruct3} can be rewritten as
\begin{equation}\label{reconstr4}
  \prod_{k=1}^{i-1} \widehat{t}(\xi_k) \cdot \widehat{T}_{\beta \alpha} (\xi_i)
  = E_i^{\alpha\beta}    \, \widehat{\tau}_s^{\alpha-\beta}\, \prod_ {k=1}^i \widehat{t}(\xi_k),  
\end{equation}
which ends the proof of \eqref{inv-pb1}.

\eqref{inv-pb2} can be proven by induction from \eqref{reconstr4}.
\qed

\section{Determinant representation for finite-size form factors}
\label{sec-ff}

In this section we consider finite-size form factors of the 8VSOS model in the cyclic case, i.e. matrix elements of local operators between Bethe eigenstates of the finite-size transfer matrix.
We recall that the space of states of the 8VSOS model corresponds to the space of functions of the height $s$ with values in the zero-weight space $\mathcal{H}[0]$, where $\mathcal{H}=V^{\otimes N}$.
Note that the action of an operator of the type $E_i^{+-}$ or $E_i^{-+}$ on a zero-weight state of $\mathcal{H}[0]$ results in a new state of $\mathcal{H}$ which is no longer of zero-weight, and therefore does not belong to the space of states of the model.
Hence, the only physical form factors of the model are those involving local operators of the type $E_i^{++}$ or $E_i^{--}$.

Let us first consider the matrix element of the operator $E_i^{--}$ at site $i$ between two Bethe eigenstates $\bra{\{u\},\omega_u }$ and $\ket{\{v\},\omega_v }$, where $\{u\}\equiv\{ u_1,\ldots , u_n\}$ and $\{ v\}\equiv\{ v_1,\ldots , v_n\}$ are solutions of Bethe equations, associated respectively to $\omega_u$ and $\omega_v$.
From the solution \eqref{inv-pb1} of the inverse problem, we get
\begin{equation}\label{inv-pbD}
    E_i^{--}    
    = \prod_{k=1}^{i-1} \widehat{t}(\xi_k) \cdot \widehat{D} (\xi_i)
        \cdot \prod_{k=i}^1\big[ \widehat{t}(\xi_k) \big]^{-1} .
\end{equation}
Hence we have
\begin{multline}
   \bra{\{u\},\omega_u }\, E_i^{--} \, \ket{\{v\},\omega_v }
   = \frac{ \prod_{k=1}^{i-1} \tau(\xi_k;\{ u \},\omega_u )}{ \prod_{k=1}^{i} \tau(\xi_k;\{ v\},\omega_v ) } \
   \frac{1}{L} \sum_{s\,\in \, \mathbf{C_{s_0}^{L}} } 
             \widetilde{\varphi}_{\omega_u} (s)\,\varphi_{\omega_v} (s-1)     \\
    \times  
         \bra{0}C(u_n;s-n) \ldots  C(u_1;s-1)\, D(\xi_i;s) \, B(v_1;s-1) \ldots B(v_n;s-n) \ket{0}. \label{ff1}
\end{multline}
 Using the commutation relation, which follows from \eqref{RTT}, of the operator $D$ with the operators $B$, we can express \eqref{ff1} in terms of the partial scalar product \eqref{result-sp}:
\begin{align}
   \bra{\{u\},\omega_u }\, E_i^{--} \, \ket{\{v\},\omega_v }
   &= \frac{ \prod_{k=1}^{i-1} \tau(\xi_k;\{ u \},\omega_u )}{ \prod_{k=1}^{i} \tau(\xi_k;\{ v\},\omega_v ) } \
   \frac{1}{L} \sum_{s\,\in \, \mathbf{C_{s_0}^{L}} } 
             \widetilde{\varphi}_{\omega_u} (s)\,\varphi_{\omega_v} (s-1)
                      \nonumber \\
   & \hspace{0.2cm} \times
        \frac{[s-n-1]\, [1]}{[s]\, [s-1]} \sum_{j=1}^n (-1)^{r\aleph} d(v_j) \frac{[s+v_j-\xi_i]}{[v_j-\xi_i]}
         \prod_{l\not= j} \frac{[v_j-v_l+1]}{[v_j-v_l]}
         \nonumber \\
   & \hspace{0.2cm} \times
         S_n(\{u\} ; \{v_\beta\}_{\beta\not= j}\cup\{\xi_i\}; s ).
         \label{ff2}
\end{align}
From  \eqref{phi-FV},\eqref{tphi-FV}, and the value \eqref{result-sp} of the partial scalar product, we get
\begin{align}
   \bra{\{u\},\omega_u }\, E_i^{--} \, \ket{\{v\},\omega_v }
   &= \frac{ \prod_{k=1}^{i-1} \tau(\xi_k;\{ u \},\omega_u )}{ \prod_{k=1}^{i} \tau(\xi_k;\{ v\},\omega_v ) } \
         \frac{1}{[\gamma]^n}\frac{\prod_{t=1}^n d(u_t)}{\prod_{k<l}[u_k-u_l][v_l-v_k]}
         \nonumber\\ 
   &\hspace{-8mm}\times        
         \sum_{j=1}^n (-1)^{r\aleph+1} d(v_j) \prod_{l=1}^n \frac{[v_j-v_l+1]}{[\xi_i-v_l]}
         \sum_{ S,\tilde{S}\subset\{1,\ldots, n\}  }  (-1)^{|S|+|\tilde{S}|}       
  \nonumber\\
 &\hspace{-8mm}\times
   \prod_{k\not\in\tilde{S}} \Bigg\{(-1)^{r\aleph} a(\hat{v}_k) \prod_{t=1}^n [u_{t}-\hat{v}_k+1] \Bigg\}
   \prod_{k\in\tilde{S}} \Bigg\{\frac{d(\hat{v}_k)}{\omega_u^{2}} \prod_{t=1}^n [u_{t}-\hat{v}_k-1] \Bigg\}
   \nonumber\\
      &\hspace{-8mm}\times
 \Bigg\{  \frac{1}{L} \sum_{s\,\in \, \mathbf{C_{s_0}^{L}} } 
 \frac{\omega_u^{-s}\omega_v^{s-1} \,[ s+v_j-\xi_i]}{ [ |u| -|v|+v_j-\xi_i+\gamma+s] }
   \frac{[\gamma+s-|S|+|\tilde{S}| ]}{[s-|S|+|\tilde{S}| ]} \Bigg\}
   \nonumber\\
     & \hspace{4.5cm} \times
         \det_n \big[ \mathcal{N}_\gamma (\{u\}; \{ \hat{v}-\delta^{S\,\tilde{S} } \}  ) \big]  ,  
         \label{ff3}
\end{align}
where $\gamma$ is arbitrary, and $\hat{v}_k=v_k$ if $k\not=j$, $\hat{v}_j=\xi_i$.
Setting $\gamma=-|u|+|v|$ and performing a change of indices in the sum over $s$, we see that, similarly as for the computation of the scalar product, we can factorize out the sum over $s$ and use the linearity of the determinant to recast the last sum over determinants into a single one:
\begin{align}
  \bra{\{u\},\omega_u }\, E_i^{--} \, \ket{\{v\},\omega_v }
   &= \frac{ \prod_{k=1}^{i-1} \tau(\xi_k;\{ u \},\omega_u )}{ \prod_{k=1}^{i} \tau(\xi_k;\{ v\},\omega_v ) } \,
          \Bigg\{ \frac{1}{L} \sum_{s\,\in \, \mathbf{C_{s_0}^{L}} } 
          \omega_u^{-s}\omega_v^{s-1} \frac{[\gamma+s ]}{[s]} \Bigg\}
         \nonumber\\
   &\times 
         \frac{\prod_{t=1}^n d(u_t)}{\prod_{k<l}[u_k-u_l][v_l-v_k]}       
         \sum_{j=1}^n (-1)^{r\aleph+1} d(v_j) \prod_{l=1}^n \frac{[v_j-v_l+1]}{[\xi_i-v_l]} 
        \nonumber\\
   & \hspace{4.5cm} \times
         \det_n \big[  \Omega_\gamma(\{u\},\omega_u;\{\hat{v}\},\omega_v)\big],   
         \label{ff4}
\end{align}
with $\Omega_\gamma$ given by \eqref{def-Omega}.
Finally, 
the remaining sum can be seen as the expansion of the determinant of the sum of two matrices, one of which being of rank 1:
\begin{align}
   \bra{\{u\},\omega_u }\, E_i^{--} \, \ket{\{v\},\omega_v }
   &= \Bigg\{ \prod_{k=1}^{i-1} \frac{\tau(\xi_k;\{ u \},\omega_u ) }{ \tau(\xi_k,\{ v \},\omega_v )}\Bigg\}  \,
          \Bigg\{  \frac{1}{L} \sum_{s\,\in \, \mathbf{C_{s_0}^{L}} } 
          \omega_u^{-s}\omega_v^{s} \frac{[\gamma+s ]}{[s]} \Bigg\} 
                 \nonumber \\
  & \times
          \frac{\prod_{t=1}^n d(u_t)}{\prod_{k<l}[u_k-u_l][v_l-v_k]}  
         \Big\{\det_n \big[  \Omega_\gamma(\{u\},\omega_u;\{v\},\omega_v) \big]
              \nonumber \\
 & 
 \ \       -
         \det_n \big[  \Omega_\gamma(\{u\},\omega_u;\{v\},\omega_v)
                              -\mathcal{P}_\gamma(\{u\},\omega_u;\{v\},\omega_v|\xi_i)\big] \Big\},   
         \label{ff5}
\end{align}
where we have defined the rank 1 matrix $ \mathcal{P}_\gamma$ as
\begin{multline}\label{mat-P}
    \big[ \mathcal{P}_\gamma(\{u\},\omega_u;\{v\},\omega_v|\xi_i) \big]_{ab} 
    =   \frac{1}{[\gamma]} \bigg\{ 
          \frac{[u_{a} - \xi_i + \gamma]}{[u_{a} - \xi_i]} 
       - \frac{\omega_v}{\omega_u} \frac{[u_{a} - \xi_i + \gamma + 1]}{[u_{a} - \xi_i + 1 ]}  \bigg\} 
         \\
        \times 
        (-1)^{r\aleph}
        a(v_{b}) \prod_{t=1}^n \bigg\{ [v_{t} - v_{b}+1] \frac{[u_{t} - \xi_i +1 ]}{[v_{t} - \xi_i + 1]}  \bigg\}  .
\end{multline}
Note that we have used the Bethe equations for $\{v\}$ to express  $ \mathcal{P}_\gamma$ as in \eqref{mat-P}.

\begin{rem}
  When $\{u\}\not=\{v\}$, one can use the orthogonality of the corresponding Bethe eigenstates and the fact that $\det_n \big[  \Omega_\gamma(\{u\},\omega_u;\{v\},\omega_v) \big]=0$ to simplify the expression \eqref{ff5}.
\end{rem}

The computation of the matrix element $\bra{ \{u \},\omega_u } E_i^{++} \ket{ \{v\},\omega_v }$ can be performed in a similar way, by using the solution of the quantum inverse problem for the operator $E_i^{++}$ in terms of the operator $\widehat{A}(\xi_i)$, and by computing the action of the latter on the state $\ket{ \{v\},\omega_v }$.
One can also notice that $E_i^{++} = 1- E_i^{--}$ and use the previous result concerning the form factor of $E_i^{--} $. 
One obtains:
\begin{multline}
   \bra{ \{u \},\omega_u } E_i^{++} \ket{ \{v\},\omega_v }
   = \Bigg\{ \prod_{k=1}^{i-1} \frac{\tau(\xi_k;\{ u \},\omega_u ) }{ \tau(\xi_k;\{ v \},\omega_v )}\Bigg\}  
          \Bigg\{ \frac{1}{L} \sum_{s\,\in \, \mathbf{C_{s_0}^{L}} } 
          \omega_u^{-s}\omega_v^{s} \frac{[\gamma+s ]}{[s]} \Bigg\}
         \\
   \times
         \frac{\prod_{t=1}^n d(u_t)}{\prod_{k<l}[u_k-u_l][v_l-v_k]}   
         \det_n \big[  \Omega_\gamma(\{u\},\omega_u;\{v\},\omega_v)
                              -\mathcal{P}_\gamma(\{u\},\omega_u;\{v\},\omega_v|\xi_i)\big].   
         \label{ff6}
\end{multline}

Combining these two results, one gets the following expression for the form factor of the operator $\sigma_i^z$ between two Bethe eigenstates:
\begin{multline}
   \bra{\{u\},\omega_u }\, \sigma_i^{z} \, \ket{\{v\},\omega_v }
   = \Bigg\{ \prod_{k=1}^{i-1} \frac{\tau(\xi_k;\{ u \},\omega_u ) }{ \tau(\xi_k;\{ v \},\omega_v )}\Bigg\}  
          \Bigg\{  \frac{1}{L} \sum_{s\,\in \, \mathbf{C_{s_0}^{L}} }
          \omega_u^{-s}\omega_v^{s} \frac{[\gamma+s ]}{[s]} \Bigg\}
         \\
   \times
         \frac{\prod_{t=1}^n d(u_t)}{\prod_{k<l}[u_k-u_l][v_l-v_k]}   
         \det_n \big[  \Omega_\gamma(\{u\},\omega_u;\{v\},\omega_v)
                              -2\mathcal{P}_\gamma(\{u\},\omega_u;\{v\},\omega_v|\xi_i)\big].   
         \label{ff-sigmaz}
\end{multline}
We recall that, in all these expressions, $\gamma=-|u|+|v|$.
Note that these formulas are very similar to those obtained in the XXZ case.

\section{Master equation representation for the two-point function}
\label{sec-master}

We now consider the problem of the computation of the two-point function, which can be tackled by summation over the corresponding form factor series.
We explain here how such a summation can be formally performed in finite volume, hence resulting into
a multiple integral representation for the two-point function, similar to the representation that was introduced in \cite{KitMST05a} in the XXZ case under the name of {\em master equation} representation.
This terminology was due to the fact that such a formula could be understood as the common result of two possible ways of computing the two-point function as a sum over elementary objects: over form factors on the one hand, or over elementary building blocks on the other hand.

We consider here a quantity which is the analog of the generating function studied in \cite{KitMST05a}.
Namely, for a generic complex number $\kappa$, we introduce the operator
\begin{equation}\label{Qkappa}
  \mathcal{Q}_{1,m}^\kappa
  = \prod_{j=1}^m \bigg(\frac{1+\kappa}{2}+\frac{1-\kappa}{2}\sigma_j^z\bigg)
  =   \prod_{j=1}^m \big( E_j^{11}+\kappa E_j^{22}\big).
\end{equation} 
The mean value of this operator in the ground state $\ket{\{ u \},\omega_u }$ of the transfer matrix,
\begin{equation}\label{generating}
  \moy{\mathcal{Q}_{1,m}^\kappa}
  = \frac{\bra{\{u\},\omega_u}\, \mathcal{Q}_{1,m}^\kappa \,\ket{\{u\},\omega_u} }{\moy{\{u\},\omega_u \,|\, \{u\} ,\omega_u} },
\end{equation}
which can be seen as a polynomial of degree $m$ in $\kappa$,
is a generating function for the two-point function of the finite-size one-dimensional quantum dynamical model whose Hamiltonian is obtained from the finite-size transfer matrix.
In the thermodynamic limit, it also gives the probability, in the CSOS model, that two sites on a same line at distance $m$ have a difference of height $\ell\le m$: the latter is given by the coefficient of $\kappa^{\frac{m-\ell}{2}}$ of \eqref{generating}\footnote{We suppose here that $m\ll L$. In fact, in this section, we are not interested in the specific properties of the root of unity case, but rather in obtaining a formula that could be extended by continuity to the case of irrational $\eta$. Therefore, when necessary, we may suppose $L$ large enough (for instance $L>N$) to avoid problems of cyclicity on the lattice.}.
 
The quantity \eqref{generating} can be evaluated by summation over form factors corresponding to a suitable basis of the space of states.
In the present case, such states can be conveniently chosen as the eigenstates of the $\kappa$-twisted transfer matrix $\widehat{t}_\kappa(u)=\widehat{A}(u)+\kappa\widehat{D}(u)$. The latter can be constructed similarly as in the untwisted case, as states of the form \eqref{state}-\eqref{phi-FV} or \eqref{tphi-FV}-\eqref{dual} with parameters $\{v\}_{_{\!\kappa}}$ satisfying the following system of $\kappa$-twisted Bethe equations:
\begin{equation}\label{Bethe-twist}
    a(v_j)  \prod_{l\ne j} \frac{[v_l-v_j+1]}{[v_l-v_j]} 
    =  (-1)^{r\aleph} \kappa\, \omega^{-2} \; d(v_j)  \prod_{l\ne j} \frac{[v_j-v_l+1]}{[v_j-v_l]} ,
    \quad j=1,\ldots n.
\end{equation}
The corresponding eigenvalues are
\begin{equation}
   \tau_\kappa (u; \{v\}_{_{\!\kappa}},\omega)
    = \omega \; a(u) \prod_{l=1}^n \frac{[v_l-u+1]}{[v_l-u]}
        +  (-1)^{r\aleph} \kappa\, \omega^{-1} \; d(u) \prod_{l=1}^n \frac{[u-v_l+1]}{[u-v_l]}.
    \label{tau-kappa}
\end{equation}
%

Indeed, from the solution \eqref{inv-pb2} of the inverse problem, the operator \eqref{Qkappa} can be expressed as follows:
\begin{equation}
   \mathcal{Q}_{1,m}^\kappa=\prod_{i=1}^m \widehat{t}_\kappa(\xi_i)\cdot\prod_{i=m}^1\big[\widehat{t}(\xi_i)\big]^{-1}.
\end{equation}
Hence, provided the eigenstates $\ket{\{v\}_{_{\!\kappa}} ,\omega_v}$ of $\widehat{t}_\kappa(u)$ form a complete basis of the space of states, one can expand the mean value \eqref{generating} as
\begin{align}
    \moy{\mathcal{Q}_{1,m}^\kappa}
  &= \sum_{ \{v\}_{_{\!\kappa}} ,\,\omega_v}
        \frac{\moy{\{u\},\omega_u\,|\,\{v\}_{_{\!\kappa}},\omega_v } \, \bra{\{v\}_{_{\!\kappa}},\omega_v } \mathcal{Q}_{1,m}^\kappa \ket{\{u\},\omega_u} }
                {\moy{\{u\},\omega_u \,|\, \{u\} ,\omega_u} \, \moy{\{v\}_{_{\!\kappa}}  ,\omega_v\,|\, \{v\}_{_{\!\kappa}}  ,\omega_v} },
      \nonumber\\
  &= \sum_{\{v\}_{_{\!\kappa}} ,\,\omega_v } \, 
        \prod_{i=1}^m \frac{\tau_\kappa(\xi_i ; \{v\}_{_{\!\kappa}},\omega_v)}{\tau(\xi_i ; \{u\},\omega_u)} \,
        \frac{\moy{\{u\},\omega_u\,|\,\{v\}_{_{\!\kappa}},\omega_v} \,  \moy{\{v\}_{_{\!\kappa}},\omega_v \, | \, \{u\},\omega_u} }
                {\moy{\{u\},\omega_u \,|\, \{u\},\omega_u } \,  \moy{\{v\}_{_{\!\kappa}},\omega_v \,|\, \{v\}_{_{\!\kappa}},\omega_v } },
                \label{average}
\end{align}
and then use the determinant representations obtained in Section~\ref{sec-sc-pdt} to express the scalar products.

In the untwisted case, for a rational parameter $\eta = \frac{1}{L}$ and for $n=N/2$, the completeness of quasi $L$-periodic Bethe eigenstates \eqref{state}-\eqref{phi-FV} with generic multiplier $\alpha$ (i.e. associated to $\omega$ such that $(-1)^{n}\omega^L=\alpha$) and generic inhomogeneity parameters $\{\xi\}$ was shown by Felder, Tarasov and Varchenko in \cite{FelTV97}, at least for any large enough odd integer $L>n$.
Their proof can easily be extended to the case of $L$-periodic $\kappa$-twisted Bethe eigenstates with arbitrary $\kappa$ (at least in a vicinity of $\kappa=0$, which is enough for the study of the polynomial quantity \eqref{generating}), and to more general rational parameters $\eta=r/L$ for $L$ odd and large enough.
This problem is discussed in Appendix~\ref{app-completeness}.
Hence, in \eqref{average}, the sum should be understood over all admissible\footnote{\label{foot-admissible}A solution $\{v\}$ of the system \eqref{Bethe-twist} is called admissible, if $\prod_{k=1}^n[v_j-\xi_k]\prod_{l=1}^n[v_j-v_l+1]\not=0,\quad j =1,...,n$.} off-diagonal\footnote{See footnote~\ref{foot-odiag}.} solutions $\{v\}_{_{\!\kappa}}$ of the $\kappa$-twisted Bethe equations \eqref{Bethe-twist} with $n=N/2$ ($\aleph=0$), associated to all possible values of $\omega_v\in\mathbb{C}$ such that $(-1)^{rn}\omega_v^L=1$.

The computation of the scalar product performed in Section~\ref{sec-sc-pdt} can easily be generalized to the $\kappa$-twisted case.
Using the same arguments as for \eqref{scal-prod}, one obtains that the scalar product between a $\kappa$-twisted Bethe eigenstate $\bra{\{v\}_{_{\!\kappa}},\omega_v}$ and a general state $\ket{\{w\},\omega_w}$ of the form \eqref{state}-\eqref{phi-FV} with $n=N/2$ can be represented as
\footnote{Here we only give the case $n=N/2$ (i.e. $\aleph=0$). In the general case, there is an additional sign in the expression of $\Omega_\gamma^{(\kappa)}$ as in \eqref{def-Omega}. } 
\begin{multline}\label{sc-kappa}
 \moy{ \{v\}_{_{\!\kappa}},\omega_v\, |\, \{w\} ,\omega_w}
 = 
     \Bigg\{ \frac{1}{L}  \sum_{s\,\in \, \mathbf{C_{s_0}^{L}} } \frac{\omega_w^{s}}{\omega_v^{s}}
     \frac{[\gamma+s]}{[s]} \Bigg\}\,
     \frac{ \prod_{t=1}^{n}   \, d(v_t)  }{ \prod_{j<k} [v_{j}-v_{k}] [w_{k}-w_{j}] } \\
     \times  \det_n \big[ \Omega_\gamma^{(\kappa)} (\{v\}_{_{\!\kappa}},\omega_v; \{ w\} ,\omega_w  ) \big]  ,
\end{multline}
with $\gamma=-|v|+|w|$, and
\begin{multline}\label{Om-kappa}
   \hspace{-2mm}
   [ \Omega_\gamma^{(\kappa)} (\{v \},\omega_v; \{ w \} ,\omega_w  ) \big]_{ij} 
   = \frac{1}{[\gamma]}\Bigg\{ \! \frac{[v_{i}-w_{j}+\gamma]}{[v_{i}-w_{j}]}
                  -\frac{\omega_w}{\omega_v}\frac{[v_{i}-w_{j}+\gamma+1]}{[v_{i}-w_{j}+1]} \! \Bigg\} 
       a(w_{j}) \!\prod_{t=1}^{n} [v_{t}-w_{j}+1]
       \\
     + \frac{1}{[\gamma]}
     \Bigg\{\! \frac{[v_{i}-w_{j}+\gamma]}{[v_{i}-w_{j}]}
                  -\frac{\omega_v}{\omega_w}\frac{[v_{i} -w_{j}+\gamma-1]}{[v_{i}-w_{j}-1]} \! \Bigg\}    
        \frac{\kappa }{\omega_{v}^{2}}\, d(w_{j})\!\prod_{t=1}^{n} [v_{t}-w_{j}-1] .
\end{multline}        
The ``square of the norm'' of a $\kappa$-twisted Bethe eigenstate is given by \eqref{gaudin} or, alternatively, in terms of the function
\begin{equation}\label{Y-kappa}
  \mathcal{Y}_{\kappa;\omega}(u;\{v\}) 
  = a(u)  \prod_{l} [v_l-u+1]
    + \kappa \omega^{-2} \, d(u)  \prod_{l} [v_l-u-1],
\end{equation}
as
\begin{equation}\label{norm-kappa}
 \moy{ \{v\}_{_{\!\kappa}},\omega_v\, |\, \{v\}_{_{\!\kappa}},\omega_v }
 = 
      \frac{ 1}{(-[0]')^n} \frac{\prod_{t=1}^n d(v_{t})}{\prod_{j\neq k} [v_j-v_k]}\cdot
     \det_n \bigg[ \frac{\partial}{\partial v_k} \mathcal{Y}_{\kappa;\omega_v}(v_j |\{v\}_{_{\!\kappa}}) \bigg].
\end{equation}

Hence, using these determinant representations for the scalar products,
one can rewrite \eqref{average} as
\begin{multline}\label{sum-Q}
  \moy{\mathcal{Q}_{1,m}^\kappa}
  =   \frac{([0]')^{2n}}{L^2} \hspace{-1mm}
       \sum_{\{v\}_{_{\!\kappa}}, \omega_v } 
       \prod_{i=1}^m \frac{\tau_\kappa(\xi_i;\{v\}_{_{\!\kappa}} ,\omega_v)}{\tau(\xi_i;\{u\},\omega_u)} \,
    \Bigg\{   \sum_{s\,\in \, \mathbf{C_{s_0}^{L}} } \hspace{-1mm}
                  \frac{\omega_v^{s}}{\omega_u^{s}} \,  \frac{[\gamma_v+s]}{[s]} \Bigg\}
    \Bigg\{   \sum_{s\,\in \, \mathbf{C_{s_0}^{L}} } \hspace{-1mm}
                  \frac{\omega_u^{s}}{\omega_v^{s}}\,  \frac{[-\gamma_v+s]}{[s]} \Bigg\}
     \\
  \times
     \frac{ \det_n \big[ \Omega_{\gamma_v} (\{u\},\omega_u ; \{ v \}_{_{\!\kappa}} ,\omega_v  ) \big]
               \cdot
               \det_n \big[ \Omega_{-\gamma_v}^{(\kappa)} (\{v\}_{_{\!\kappa}},\omega_v ; \{ u \} ,\omega_u  ) \big] }
             { \det_n \big[ \frac{\partial}{\partial u_k} \mathcal{Y}_{1;\omega_u}(u_j;\{u\})\big]
               \cdot
               \det_n \big[ \frac{\partial}{\partial v_k} \mathcal{Y}_{\kappa;\omega_v}(v_j;\{v\}_{_{\!\kappa}})\big]} ,
\end{multline}  
in which we have set $\gamma_v=|v|-|u|$.
As in \cite{KitMST05a}, the sum over admissible off-diagonal solutions of the $\kappa$-twisted Bethe equations in \eqref{sum-Q} can be reexpressed as a multiple integral, and one obtains the following representation ({\em master equation}) for the generating function \eqref{generating}:
\begin{multline}\label{master-eq}
 \moy{\mathcal{Q}_{1,m}^\kappa} 
 =\frac{1}{n!} \frac{([0]')^{2n}}{L^2} \sum_\omega \oint\limits_{\Gamma(\{v\}_{_{\!\kappa}})} \frac{d^n z}{(2i\pi)^n} \,
        \prod_{i=1}^m \frac{\tau_\kappa(\xi_i;\{z\},\omega )}{\tau(\xi_i;\{u\},\omega_u)} \,
    \Bigg\{   \sum_{s\,\in \, \mathbf{C_{s_0}^{L}} }
                        \frac{ \omega^{s}}{\omega_u^{s}}\,
   \frac{[\gamma_z+s]}{[s]} \Bigg\}^{\! 2}  
          \\
  \times   \frac{ \det_n \big[ \Omega_{\gamma_z} (\{u\},\omega_u; \{ z \} ,\omega ) \big]
               \cdot
               \det_n \big[ \Omega_{-\gamma_z}^{(\kappa)} (\{z\},\omega; \{ u \} ,\omega_u ) \big] }
             { \det_n \big[ \frac{\partial}{\partial u_k} \mathcal{Y}_{1;\omega_u}(u_j;\{u\})\big]
               \cdot
               \prod_{j=1}^n  \mathcal{Y}_{\kappa;\omega}(z_j;\{z\}) }          .
\end{multline}
The sum in \eqref{master-eq} is taken over all $\omega\in\mathbb{C}$ such that $(-1)^{rn}\omega^L=1$, and the integration contour is such that it surrounds (with index 1) all poles corresponding to solutions $\{v\}_{_{\!\kappa}}$ with $n=N/2$ of the $\kappa$-twisted Bethe equations associated to $\omega$.
The other poles of the integrand, which lie outside the contour, are poles at $z_\ell=\xi_i$, $i=1,\ldots,m$ (issued from the factors $\tau_\kappa(\xi_i;\{z\},\omega)$), and poles at $z_\ell=u_j$, $j=1,\ldots,n$ (contained in $\det_n \big[ \Omega_{-\gamma_z}^{(\kappa)} (\{z\},\omega; \{ u \} ,\omega_u  ) \big] $).
By considering these poles, one can, as in the XXZ case \cite{KitMST05a}, rewrite the above integral in the alternative form
\begin{multline}
  \moy{\mathcal{Q}_{1,m}^\kappa}
  =\frac{1}{n!} \frac{([0]')^{n}}{L^2} \sum_\omega
     \oint\limits_{\Gamma(\{\xi\}) \cup \Gamma(\{ u \})} \frac{d^n z}{(2i\pi)^n} 
        \prod_{i=1}^m \frac{\tau_\kappa(\xi_i;\{z\} ,\omega )}{\tau(\xi_i;\{u\},\omega_u)}\,
    \Bigg\{   \sum_{s\,\in \, \mathbf{C_{s_0}^{L}} }
                         \frac{\omega^{s}}{\omega_u^{s}}\, \frac{[\gamma_z+s]}{[s]} \Bigg\}^{\! 2}  
          \\
  \times   \frac{ \det_n \big[ \Omega_{\gamma_z} (\{u\},\omega_u; \{ z \} ,\omega ) \big]
               \cdot
               \det_n \big[ \Omega_{-\gamma_z}^{(\kappa)} (\{z\},\omega; \{ u \} ,\omega_u  ) \big] }
             { \det_n \big[ \Omega_0 (\{u\},\omega_u; \{ u \} ,\omega_u  )  \big]
               \cdot
               \prod_{j=1}^n  \mathcal{Y}_{\kappa;\omega}(z_j;\{z\}) }  .
\end{multline}

\section{Conclusion}
\label{sec-concl}

In this paper we have generalized the algebraic Bethe Ansatz approach to correlation functions to the dynamical case, i.e. to the case of an $R$-matrix depending on a dynamical parameter and satisfying the dynamical Yang-Baxter equation.
In this framework, we have obtained finite-size representations for scalar products of Bethe states, form factors of local operators and two-point correlation function. These results hence initiate the study of correlation functions of the (cyclic) 8VSOS model within the framework of algebraic Bethe Ansatz. Applications of these formulas at the thermodynamic limit will be considered in a further publication.

As we have seen, one of the difficulty of the ABA approach to correlation functions is that it relies on the existence of a compact and manageable expression, preferably as a single determinant, for the scalar products of Bethe states: in practice, such a representation may not be easy to obtain, and it is even possible that it does not exist for all models, at least in the simplest expected form of a single determinant (see for instance the recent preprints \cite{BelPRS12a,BelPRS12b} concerning $SU(3)$-invariant models).
In the case of general SOS models, in fact, it seems that we can only obtain representations of height-dependent (partial) scalar products \eqref{Sn} as sums of determinants. As in the case of the partition function with domain wall boundary conditions \cite{Ros09,PakRS08}, this is due to the shifts undergone by the dynamical parameter. 
However, as mentioned in Section~\ref{sec-sc-pdt}, it is not very surprising that  shifts of the dynamical parameter 
prevent us from representing the height-dependent partial scalar product (and hence the partition function) as a single determinant, since such shifts also prevent the corresponding states, i.e. states of $\mathcal{H}[0]$ of the form \eqref{partial-state}, from directly being Bethe states.
In fact, when one considers true Bethe states as functions of the dynamical parameter, and their natural scalar products which, in the cyclic case, can be expressed as a finite sum over all values of the dynamical parameter, one obtains for the latter a single determinant representation very similar to its XXZ analog.
It follows in particular that the norm of a Bethe eigenstate can be expressed as a single determinant. As we have seen, this opens the way to the computation of form factors and correlation functions.

As a final remark, we would like to mention an interesting and promising alternative method: Sklyanin's separation of variables (SOV) method \cite{Skl90,Skl92}, which has been recently developed in the direction of the computation of form factors \cite{GroMN12,Nic12a,Nic12b}. The advantage of this approach is that it enables one to free oneself from the main difficulty of the ABA approach to correlation functions, namely the fact that, depending on the model we consider, we have a priori no guaranty that a convenient expression for the scalar products does exist and --- should it exist --- no model-independent constructive method to derive it.  In fact, in the SOV approach, it seems that scalar products of separates states can automatically be expressed as a determinant. Let us note however that, at least at the present stage of the art, the determinant representation obtained from ABA still presents some advantages, the main one being probably that it is more appropriate than its SOV counterpart for taking the thermodynamic limit.

\section*{Acknowledgements}

We thank G. Filali, N. Kitanine, G. Niccoli and J.M. Maillet for their interest in this work.
V. T. would also like to thank M. Jimbo and J. Shiraishi for an old discussion at the origin of this work.

V. T. is supported by CNRS.
We also acknowledge  the support from the ANR grant DIADEMS 10 BLAN 012004.
V. T. would  like to thank LPTHE (Paris VI University) for hospitality.

\appendix

\section{Definition and useful properties of the function $u\mapsto [u]$}
\label{ap-theta}

Let us set $p=e^{2\pi i\tau}$, with $\Im{\tau} > 0$. 
Let $\theta_1(z;\tau)$ be the Theta function of quasi-periods 1 and $\tau$ and nome $p^{1/2}=e^{i \pi \tau}$:
\begin{align}
  \theta_1(z;\tau) &= 2 p^{1/8} \sin \pi z \prod_{n=1}^\infty (1-2 p^{n} \cos 2\pi z + p^{2n}) (1-p^{n})
                                         \nonumber\\
                              &= -i \sum_{n=-\infty}^\infty (-1)^n p^{\frac{1}{2}(n+\frac{1}{2})^2} e^{(2n+1)i\pi z}.
                              \label{def-theta}
\end{align}
In the generic SOS model (i.e. for $\eta$ generic), the elements \eqref{bc1}-\eqref{bc2} of the dynamical $R$-matrix are usually defined in terms of the function
\begin{equation}\label{ord-theta}
  [u]= \theta_1(\eta u; \tau),
\end{equation}
which is odd, entire, and satisfies the quasi-periodicity properties 
 \begin{equation}\label{quasi-p1}
    [u+1/\eta]=-[u], \qquad [u+\tau/\eta]=- e^{-i\pi\tau} e^{-2i\pi\eta u} [u] .
 \end{equation}
 This function also satisfies the following useful properties:
 \begin{enumerate}
 \item Addition formula:
 \begin{equation}\label{id1}
   [x+u] [x-u] [y+v] [y-v] -[x+v] [x-v] [y+u] [y-u] = [x+y] [x-y] [u+v] [u-v].
 \end{equation}
\item Let $t\in\mathbb{C}$ and $f$ be of the form
 \begin{equation}\label{ratio-theta}
    f(x)=\sum_j C^{(j)} \frac{[x-a_1^{(j)}] \cdots [x-a_{\ell +m}^{(j)}]}{[x-b_1^{(j)}] \cdots [x-b_\ell^{(j)}]},
 \end{equation}
 where $a_i^{(j)}, b_i^{(j)}, C^{(j)} \in\mathbb{C}$ such that
 $a_1^{(j)}+\cdots + a_{\ell+m}^{(j)} - b_1^{(j)} - \cdots - b_\ell^{(j)} =t\mod 1/\eta, \ \forall j$.
 Then $f$ satisfies the quasi-periodicity properties
 \begin{equation}\label{quasi-p-f}
   f(x+1/\eta) = (-1)^m f(x), \qquad f(x+\tau/\eta)=(-1)^m e^{2\pi i\eta t-i\pi m(2\eta x+\tau) }f(x).
 \end{equation}
Moreover, if $f$ is entire (i.e. if the singularities at $x=b_i^{(j)}$ are all removable), then it 
can be factored as
 \begin{equation}\label{theta-ordre-r}
   f(x)=C [x-a_1] \ldots [x-a_m],
 \end{equation}
where $C,a_1,\ldots, a_m$ are complex numbers such that $a_1+\cdots + a_m=t$.
%
\item Frobenius' determinant evaluation:
 \begin{equation}\label{Frob-det}
   \det_{1\le i,j \le n} \left( \frac{[x_i-y_j+t]}{[x_i-y_j]} \right) 
   =  [t]^{n-1} [ |x|-|y|+t] \frac{\prod_{1\le i<j\le n} [x_i-x_j][y_j-y_i]}
                                              {\prod_{i,j=1}^n [x_i-y_j]} ,
\end{equation}
where $|x|=x_1+\dots+x_n$, $|y|=y_1+\dots + y_n$.
\item For any positive integer $L$, one has the following identity:
\begin{equation}\label{sum-L}
  \frac{[u+\gamma]\,[0]'}{[u]\,[\gamma]}
  =\sum_{k=0}^{L-1} e^{2\pi i\eta k u}  \frac{[Lu+\gamma+k\frac{\tau}{\eta}]_{_L}\,[0]_{_L}'}{[Lu]_{_L}\,[\gamma+k\frac{\tau}{\eta}]_{_L}},
\end{equation}
where $[u]_{_L}=\theta_1(\eta u; L\tau)$.
\eqref{sum-L} can be shown by relabeling the  sum ($k\rightarrow k+Lj$) in the Fourier series expansion:
\begin{equation}
  \frac{[u+\gamma]\,[0]'}{[u]\,[\gamma]}=-2i\pi\eta\sum_{k=-\infty}^{+\infty} \frac{e^{2\pi i\eta k u}}{1-e^{2\pi i \eta\gamma} p^k}.
\end{equation}
\end{enumerate}

In the cyclic case, when $L\eta=r_1+r_2\tau$ with $L,r_1,r_2$ integers, it is convenient, following Baxter \cite{Bax73a}\footnote{Equation (9) of paper I of \cite{Bax73a} corresponds in fact to $L\eta=2r_1+ 2 r_2\tau$. Here we get rid of factor 2 by allowing a sign in the $L$-periodicity \eqref{per-mod2} of the function $u\mapsto [u]$.}, to re-define the function $u\mapsto [u]$ as
 \begin{equation}\label{modified-theta}
   [u]= \theta_1(\eta u; \tau) \, e^{i\pi r_2 \frac{\eta}{L} u^2}.
 \end{equation}
The modified function \eqref{modified-theta} is odd, entire, and satisfies the quasi-periodicity properties:
 \begin{equation}\label{per-mod}
    [u+1/\eta]=- e^{i\pi \frac{r_2}{L}(2u+\frac{1}{\eta})} \, [u], \qquad
    [u+\tau/\eta]=- e^{-i\pi\frac{r_1}{L}(2u+\frac{\tau}{\eta})} \, [u].
 \end{equation}
It is also quasi-periodic of period $L$ with a mere sign:
\begin{equation}\label{per-mod2}
  [u+L] =(-1)^{r_1+r_2+r_1 r_2}\, [u], 
\end{equation}
so that a ratio of two such functions is periodic of period $L$.
The modified function \eqref{modified-theta} still satisfies the properties 1. and 3. above. If moreover $f$ is of the form \eqref{ratio-theta} with
$a_1^{(j)}+\cdots + a_{\ell+m}^{(j)} - b_1^{(j)} - \cdots - b_\ell^{(j)} =t\mod L, \ \forall j$,
then $f$ satisfies the quasi-periodicity properties
\begin{equation*} 
   f(x+1/\eta) = (-1)^m e^{i\pi\frac{r_2}{L} (2mx-2t+m\frac{1}{\eta})} f(x), \quad
   f(x+\tau/\eta)=(-1)^m e^{i\pi \frac{r_1}{L}(2mx-2t+m\frac{\tau}{\eta}) } f(x),
\end{equation*}
and can be factored as in \eqref{theta-ordre-r}.
In that case, it is enough to find $m$ independent zeroes to prove that $f$ vanishes identically.

In fact, as noticed in \cite{FabM07}, the modified function \eqref{modified-theta} is nothing else but a usual theta function of the form \eqref{ord-theta} with different quasi-periods. This can easily be seen from the transformation property of the theta function \eqref{def-theta} under the action of the modular group:
\begin{equation}
    e^{i\pi\frac{c z^2}{c\tau + d}} \, \theta_1(z;\tau) \propto \theta_1\Big(\frac{z}{c\tau+d};\frac{a\tau+b}{c\tau+d}\Big), 
    \qquad\text{for any }
    \begin{pmatrix}
      a & b \\
      c & d
    \end{pmatrix}
    \in SL(2,\mathbb{Z}).
\end{equation} 
Hence the modified function \eqref{modified-theta} is simply proportional to
\begin{equation}\label{m1-m2}
   [u] \propto \theta_1( \eta' u ;  \tau'), \qquad
   \text{with\quad} \eta'=\frac{r}{L} \quad \text{and\quad} \tau'=\frac{b+a\tau}{\widetilde{r}_1+\widetilde{r}_2\tau}.
\end{equation}
In this expression, $r$ is the greatest common divisor of $r_1$ and $r_2$, and the integers $a, b, \widetilde{r}_1,\widetilde{r}_2$ are such that $r_1= r\widetilde{r}_1$, $r_2= r\widetilde{r}_2$ and $a \widetilde{r}_1- b\widetilde{r}_2=1$.
Hence all the properties of \eqref{ord-theta} can trivially be extended to \eqref{modified-theta} with the mere replacement of $\eta$ by $\eta'$ and $\tau$ by $\tau'$, and the study of the case $L\eta =r_1+r_2\tau$ is therefore equivalent to the simplest $\eta=r/L$ case.

\section{Partition function of the 8VSOS model with domain wall boundary conditions}
\label{app-part-fct}

In this appendix, we recall the result of \cite{Ros09} concerning the representation of the partition function of the 8VSOS model with domain wall boundary conditions as a sum of determinants, useful for the computation of scalar products.
For a inhomogeneous square lattice of size $N\times N$ with parameters $u_1,\ldots,u_N,\xi_1,\ldots, \xi_N$, this partition function can be expressed in terms of the entries of the monodromy matrix \eqref{monodromy} (with inhomogeneity parameters $\xi_1,\ldots, \xi_N$)  as
\begin{equation}\label{part-fct}
   Z_N(u_1,\ldots,u_N;\xi_1,\ldots,\xi_N;s)
   =\bra{0}  C   (u_{N};s)\, C   (u_{N-1};s+1) \ldots C  (u_{1};s+(N-1) ) \ket{\bar{0}},
\end{equation}
where $\ket{\bar{0}}=\e_-\otimes\dots\otimes \e_-$.
It can easily be shown that this quantity is uniquely determined by the following properties \cite{Ros09,PakRS08}:
\begin{enumerate}
  \item $\displaystyle{  Z_1(u,\xi;s) =\frac{[1][s-u+\xi]}{[s] [u-\xi+1]} }$;
  \item $Z_N(u_1,\ldots,u_N;\xi_1,\ldots,\xi_N;s)$ is symmetric in $u_{1},\ldots,u_{N} $ and in $\xi_{1},\ldots,\xi_{N}$ separately;
  \item $\prod_{\alpha,\ell=1}^N [u_{\alpha}-\xi_{\ell}+1]\, Z_N(u_1,\ldots,u_N;\xi_1,\ldots,\xi_N;s)$ is a holomorphic function of each $u_\alpha$ and each $\xi_\ell$; 
  \item it satisfies the recursion relation:
  \begin{multline}
    \frac{[u_{N}-\xi_{N}+1]}{[1]} \; Z_N (u_1,\ldots,u_N;\xi_1,\ldots,\xi_N;s) \big|_{u_{N}=\xi_{N}-1}
    =\frac{[s+N]}{[s+N-1]}
        \\
        \times
 \prod_{j=1}^{N-1}\frac{[u_{j}-\xi_{N}]}{[u_{j}-\xi_{N}+1]}
       \prod_{k=1}^{N-1} \frac{[\xi_{k}-\xi_{N}+1]}{[\xi_{k}-\xi_{N}]}       Z_{N-1}(u_{1},\ldots, u_{N-1};\xi_{1},\ldots,\xi_{N-1};s);\!\!
  \end{multline}
  note that we have also
  \begin{equation}
   Z_N (u_1,\ldots,u_N;\xi_1,\ldots,\xi_N;s)\big|_{u_{N}=\xi_{N}}
   =Z_{N-1}(u_{1},\ldots, u_{N-1};\xi_{1},\ldots,\xi_{N-1};s+1);\!
  \end{equation}
  \item $Z_N=f(u_{1})$ as a function of $u_1$ satisfies the quasi-periodicity properties:
  \begin{equation}
    f(u_{1}+1/\eta)= e^{-2i\pi\frac{r_2}{L}(s+N)} f(u_{1}), \qquad
    f(u_{1}+\tau/\eta)= e^{-2i\pi\frac{r_1}{L}(s+N)} f(u_{1}).
  \end{equation}
\end{enumerate}
In \cite{Ros09,PakRS08}, a representation of the partition function as a sum over the permutations of the set $\{\xi\}$ was proposed. In \cite{Ros09}, it was shown that this representation could be reduced to the following sum of determinants:
\begin{multline}\label{Z-1}
Z_N(\{ u\};\{\xi \};s )
=\frac{[s+N]}{[\gamma]^N\, [ |u | - | \xi | + \gamma +s +N ]} \,
\frac{\prod_{\alpha, \ell =1}^N [ u_{\alpha} -\xi_{\ell} ]}
        {\prod_{j<k} [u_{j}-u_{k} ] \, [\xi_{k}-\xi_{j} ] } \\
        \times
  \sum_{S\subset\{1,\ldots, N\} } (-1)^{|S|} \, \frac{[\gamma+s+N-|S| ]}{[s+N-|S| ]} \;
       \det_N \big[ \mathcal{N}_\gamma(\{u+\delta^S\};\{\xi \})\big].
\end{multline}
In this expression, $\gamma$ is a generic parameter (the left-hand side being independent of $\gamma$), $\mathcal{N}_\gamma$ is given by \eqref{det-N}, $|u |=u_1+\dots+u_N$, $|\xi |=\xi_1+\dots+\xi_N$, the sum runs over all subsets $S$ of $\{1,\ldots, N\}$, with $|S|$ being the cardinality of $S$ and $\{u+\delta^S\}=\{u_j-\delta_j^S\}_{1\le j \le N}$ with
\begin{equation}\label{deltaS}
  \delta_{j}^S =
  \begin{cases}
       1 & \text{if } j\in S, \\
       0 & \text{if } j\notin S.
  \end{cases}
\end{equation}
Formula \eqref{Z-1} was proved in \cite{Ros09} by induction over $N$. It is easy to see that, following the lines of the proof of \cite{Ros09}, one can also obtain the equivalent formula:
\begin{multline}\label{Z-2}
Z_N(\{ u\};\{\xi \};s )
=\frac{[s+N]}{[\gamma]^N\, [ |u | - | \xi | + \gamma +s +N ]} \,
\frac{\prod_{\alpha, \ell =1}^N [ u_{\alpha} -\xi_{\ell} ]}
        {\prod_{j<k} [u_{j}-u_{k} ] \, [\xi_{k}-\xi_{j} ] } \\
        \times
  \sum_{S\subset\{1,\ldots, N\} } (-1)^{|S|} \, \frac{[\gamma+s+N-|S| ]}{[s+N-|S| ]} \;
        \det_N \big[ \mathcal{N}_\gamma(\{u\};\{\xi-\delta^S\})\big].
\end{multline}

\begin{rem}\label{rem-pole-part}
The partition function $Z_N(\{ u\};\{\xi \};s )$ has no pole at $ |u | - | \xi | + \gamma +s +N=0$ (it is holomorphic except at the points $u_\alpha-\xi_\ell+1=0$), which means that the sum of determinants in the second line of \eqref{Z-1} or \eqref{Z-2} has a zero at this point.
\end{rem}

\section{Solution of the recursion relation \eqref{induction} for the scalar product}
\label{app-sc}

We show here that the solution of the recursion relation \eqref{induction},\eqref{coeff-rec}, with the initial condition \eqref{initial}, is given by the following expression:
 \begin{multline}
    G^{(k)}_{\ell_{k+1},\ldots, \ell_n} ( \{u\}; \{v_{1}, \ldots, v_{k} \} ; s)
  = \frac{ [s-k]}{[\gamma]^n\,  [ |u | -|\xi_{\ell} |_{n-k}-|v |_{k}+\gamma+s]}
    \\
 \times
  \prod_{j=0}^{k-1}\frac{[s-j]}{[s+N-2n+j]}
  \prod_{l=k+1}^n \left\{ \prod_{\substack{j=1 \\ j\ne \ell_l}}^N\frac{[\xi_{\ell_l}-\xi_j+1]}{[\xi_{\ell_l}-\xi_j]}
                                      \cdot \prod_{j=1}^{k}\frac{1}{[\xi_{\ell_l}-v_{j}]}  
                                      \cdot \frac{\prod_{j=1}^n [u_{j}-\xi_{\ell_l}+1]}
                                                       {\prod_{\substack{j=k+1 \\ j\ne l}}^n [\xi_{\ell_j}-\xi_{\ell_l}+1]}
                            \right\}    \\
  \times 
 \frac{\prod_{j=1}^{k} d(v_{j})  \prod_{j=1}^n d(u_j)\,\prod_{k< j< l} [\xi_{\ell_j}-\xi_{\ell_l}]}
        {\prod_{j<l} [u_{j}-u_{l}] \prod_{j<l \le k} [v_{l}-v_{j}] }  
      \sum_{\substack{S\subset\{1,\ldots, n\} \\ \tilde{S}\subset\{1,\ldots, k\} } } (-1)^{|S|+|\tilde{S}|} \,
      \frac{[\gamma+s-|S|+|\tilde{S}| ]}{[s-|S|+|\tilde{S}| ]}
     \\
  \times
       \prod_{j\not\in\tilde{S}} \Bigg\{
  \frac{a(v_{j})}{d(v_{j}) }\, \prod_{l=1}^n [u_l-v_{j}+1] \Bigg\}
   \prod_{j\in\tilde{S}} \Bigg\{(-1)^{r\aleph}\omega_u^{-2} \prod_{l=1}^n [u_l-v_{j}-1] \Bigg\}
   \\
   \times
         \det_n \big[ \mathcal{N}_\gamma (\{u\};v_{1}^{S\,\tilde{S} },\ldots, v_{k}^{S\,\tilde{S} },
                \xi_{\ell_{k+1}}^S,\ldots,\xi_{\ell_n}^S  ) \big]  . 
         \label{rec-Gk}  
 \end{multline}
Here we have used the notations of Proposition~\ref{prop-sc} with $v_j^{S\,\tilde{S} }=v_j-\delta_j^{S\,\tilde{S} }$, $\xi_{\ell_j}^S=\xi_{\ell_j}-\delta_j^S$ with $\delta_j^S$ given by \eqref{deltaS}, $|v |_{k}=v_1+\dots + v_k$, $|\xi_{\ell} |_{n-k}=\xi_{k+1}+\dots+\xi_n$.
 
Let us first remark that  \eqref{rec-Gk} provides the correct expression for $G^{(0)}_{\ell_{1},\ldots, \ell_n}$: this follows from \eqref{initial} and from the representation \eqref{Z-2} for the partition function with domain wall boundary conditions.

Supposing that \eqref{rec-Gk} is valid at level $k-1$, we now want to prove that it is also valid at level $k$. 
From the recursion relation~\eqref{induction}, \eqref{coeff-rec}, we obtain
\begin{multline}
   G^{(k)}_{\ell_{k+1},\ldots, \ell_n} ( \{u\}; \{ v_{1}, \ldots, v_{k}\} ; s)
  = \frac{ 1  }{[\gamma]^n   }  
      \prod_{j=0}^{k}\frac{[s-j]}{[s+N-2n+j]} 
     \\
    \times
   \prod_{l=k+1}^n \left\{ \prod_{\substack{j=1 \\ j\ne \ell_l}}^N\frac{[\xi_{\ell_l}-\xi_j+1]}{[\xi_{\ell_l}-\xi_j]}
                                      \cdot\frac{[\xi_{\ell_l}-v_k-1]}{\prod_{j=1}^{k}[\xi_{\ell_l}-v_{j}]}  
                                      \cdot \frac{\prod_{j=1}^n [u_{j}-\xi_{\ell_l}+1]}
                                                       {\prod_{\substack{j=k+1 \\ j\ne l}}^n [\xi_{\ell_j}-\xi_{\ell_l}+1]} 
                                \right\}
     \\                    
   \times 
     \frac{\prod_{j=1}^{k} d(v_j)  \prod_{j=1}^n d(u_j)\,\prod_{k< j< l} [\xi_{\ell_j}-\xi_{\ell_l}]}
        {\prod_{j<l} [u_{j}-u_{l}] \prod_{j<l < k} [v_{l}-v_{j}] }  \,
        \prod_{j=1}^N\frac{1}{[v_{k}-\xi_j]}
        \cdot f^{(k)}(v_{k}),
  \end{multline}
  with
  \begin{multline}\label{fp}
   f^{(k)}(v_k)= \prod_{j=1}^N [v_k-\xi_j]
       \sum_{\ell_k=1}^N \frac{[1]\, [s+N-2n+k+v_k-\xi_{\ell_k}]}
          {[v_{k}-\xi_{\ell_k}] \, [ |u| -|\xi_{\ell}|_{n-k+1}-|v|_{k-1}+\gamma+s]}
          \prod_{\substack{j=1 \\ j\ne \ell_k}}^N \!\! \frac{[\xi_{\ell_k}-\xi_j+1]}{[\xi_{\ell_k}-\xi_j]}
         \\
         \times
         \frac{\prod_{j=1}^n [u_{j}-\xi_{\ell_k}+1]}
                 {\prod_{j=k+1}^n [\xi_{\ell_j}-\xi_{\ell_k}-1]\cdot \prod_{j=1}^{k-1} [\xi_{\ell_k}-v_{j}]}
        \sum_{\substack{S\subset\{1,\ldots, n\} \\ \tilde{S}\subset\{1,\ldots, k-1\} } } (-1)^{|S|+|\tilde{S}|} \, 
        \frac{[\gamma+s-|S|+|\tilde{S}| ]}{[s-|S|+|\tilde{S}| ]}
          \\
  \times
    \prod_{\substack{j=1 \\ j\not\in\tilde{S}} }^{k-1} \Bigg\{
  \frac{a(v_{j})}{d(v_{j}) }\, \prod_{l=1}^n [u_l-v_{j}+1] \Bigg\}
   \prod_{j\in\tilde{S}} \Bigg\{(-1)^{r\aleph}\omega_u^{-2} \prod_{l=1}^n [u_l-v_{j}-1] \Bigg\}
   \\
   \times
        \det_n \big[ \mathcal{N}_\gamma (\{u\};v_{1}^{S\,\tilde{S} },\ldots, v_{k-1}^{S\,\tilde{S} },
                \xi_{\ell_{k}}^S,\ldots,\xi_{\ell_n}^S  ) \big].
\end{multline}
Note that the sum in \eqref{fp} (which corresponds to the sum in \eqref{induction}) has been extended to all values of $\ell_k$: indeed the sum over determinants,
which is antisymmetric by exchange of two different $\xi_\ell$, vanishes for $\ell_k=\ell_j$, $j=k+1,\ldots,n$. 
The function \eqref{fp} of $v_{k}$ is an entire function of the form \eqref{ratio-theta} with $m=N$ and $t=\sum_{j=1}^N\xi_j-s-N+2n-k$. It can also be written as
\begin{multline}\label{fpbis}
 f^{(k)}(v_k)= \prod_{j=1}^N [v_k-\xi_j]
       \sum_{\ell_k=1}^N \frac{[1]\, [s+N-2n+k+v_k-\xi_{\ell_k}]}
          {[v_{k}-\xi_{\ell_k}] \, [ |u| -|\xi_{\ell}|_{n-k+1}-|v|_{k-1}+\gamma+s]}
          \prod_{\substack{j=1 \\ j\ne \ell_k}}^N \!\! \frac{[\xi_{\ell_k}-\xi_j+1]}{[\xi_{\ell_k}-\xi_j]}
         \\
   \times
         \frac{ 1 }
                 {\prod_{j=k+1}^n [\xi_{\ell_j}-\xi_{\ell_k}-1]\cdot \prod_{j=1}^{k-1} [\xi_{\ell_k}-v_{j}]}
        \sum_{\substack{S\subset\{1,\ldots, n\}\setminus\{k\} \\ \tilde{S}\subset\{1,\ldots, k-1\} } } (-1)^{|S|+|\tilde{S}|}
        \\
 \times
    \prod_{\substack{j=1 \\ j\not\in\tilde{S}} }^{k-1} \Bigg\{
  \frac{a(v_{j})}{d(v_{j}) }\, \prod_{l=1}^n [u_l-v_{j}+1] \Bigg\}
   \prod_{j\in\tilde{S}} \Bigg\{(-1)^{r\aleph}\omega_u^{-2} \prod_{l=1}^n [u_l-v_{j}-1] \Bigg\}
       \\
 \times
         \det_n \big[ \hat{\mathcal{N}}^{(k)}_\gamma (\{u\};v_{1}^{S\,\tilde{S} },\ldots, v_{k-1}^{S\,\tilde{S} },
                \xi_{\ell_{k}},
                \xi_{\ell_{k+1}}^S,\ldots,\xi_{\ell_n}^S   ) \big] ,                           
\end{multline}
with $\big[ \hat{\mathcal{N}}^{(k)}_\gamma \big] _{ij}
                =\big[ {\mathcal{N}}_\gamma \big]_{ij}$ for $j\ne k$ and 
\begin{multline*}
   \big[ \hat{\mathcal{N}}^{(k)}_\gamma (\{u\};v_{1}^{S\,\tilde{S} },\ldots, v_{k-1}^{S\,\tilde{S} },
                \xi_{\ell_{k}},
                \xi_{\ell_{k+1}}^S,\ldots,\xi_{\ell_n}^S   ) \big]_{ik}
                = \prod_{j=1}^n [u_{j}-\xi_{\ell_k}+1]
                \\
            \times
            \bigg\{\frac{[\gamma+s-|S|+|\tilde{S}| ]}{[s-|S|+|\tilde{S}| ]}
              \frac{[u_i-\xi_{\ell_k}+\gamma]}{[u_{i}-\xi_{\ell_k}]}  
          -     \frac{[\gamma+s-|S|+|\tilde{S}| -1]}{[s-|S|+|\tilde{S}|-1 ]}
              \frac{[u_{i}-\xi_{\ell_k}+\gamma+1]}{[u_{i}-\xi_{\ell_k}+1]} \bigg\}
              \\
        =-\prod_{\substack{j=1\\ j\ne i} }^n[u_j-\xi_{\ell_k}+1]   \,  
        \frac{[1] [\gamma]  [u_i-\xi_{\ell_k}+s-|S|+|\tilde{S}|+\gamma] [  u_{i}-\xi_{\ell_k}-s+|S|-|\tilde{S}|  +1]   }
               {[s-|S|+|\tilde{S}| ] [s-|S|+|\tilde{S}|-1] [u_{i}-\xi_{\ell_k}] }, 
\end{multline*}
in which we have used \eqref{id1}.
This function is equal to
\begin{multline}
 f^{(k)}(v_k)=\frac{[s+N-2n+k] }{ [ |u| -|\xi_{\ell}|_{n-k}-|v|_k+\gamma+s] }
 \frac{1}{\prod_{j=k+1}^n [\xi_{\ell_j}-v_k-1] \cdot \prod_{j=1}^{k-1}[v_k-v_j]}
  \\
  \times
  \sum_{\substack{S\subset\{1,\ldots, n\}\setminus\{k\} \\ \tilde{S}\subset\{1,\ldots, k-1\} } } 
        \hspace{-2mm}
       (-1)^{|S|+|\tilde{S}|} \,
       \prod_{\substack{ j=1 \\ j\not\in\tilde{S}}}^{k-1} \Bigg\{
  \frac{a(v_j)}{d(v_j) } \prod_{l=1}^n [u_l-v_j+1] \Bigg\}
   \prod_{j\in\tilde{S}} \Bigg\{(-1)^{r\aleph}\omega_u^{-2} \prod_{l=1}^n [u_l-v_j-1] \Bigg\}
   \\
  \times
   \Bigg\{
   \prod_{j=1}^N [v_k-\xi_j+1]\,
         \det_n \big[ \hat{\mathcal{N}}^{(k)}_\gamma (\{u\};v_1^{S\,\tilde{S} },\ldots, v_{k-1}^{S\,\tilde{S} }, v_k,\xi_{\ell_{k+1}}^S,\ldots,\xi_{\ell_n}^S  ) \big] 
   \\
   -(-1)^{r\aleph}\omega_u^{-2} \prod_{j=1}^N [v_k-\xi_j] \,
        \det_n \big[ \check{\mathcal{N}}^{(k)}_\gamma(\{u\};v_1^{S\,\tilde{S} },\ldots, v_{k-1}^{S\,\tilde{S} }, v_k,\xi_{\ell_{k+1}}^S,\ldots,\xi_{\ell_n}^S  ) \big]    \Bigg\}
      \displaybreak[0]   \\
   +(-1)^{r\aleph}\omega_u^{-2}\prod_{j=1}^N [v_k-\xi_j]
      \sum_{q=k+1}^n \frac{[s+N-2n+v_k-\xi_{\ell_q}+k+1] }{ [ |u| -|\xi_{\ell}|_{n-k}-|v|_{k-1}-\xi_{\ell_q}+\gamma+s+1] }
   \frac{1}{[\xi_{\ell_q}-v_k-1]}
   \\
    \times
     \prod_{\substack{j=k+1 \\ j\ne q}}^n   \frac{1}{[\xi_{\ell_j}-\xi_{\ell_q}]} 
     \prod_{j=1}^{k-1}\frac{1}{[\xi_{\ell_q}-v_{j}-1]}
      \sum_{\substack{S\subset\{1,\ldots, n\}\setminus\{k\} \\ \tilde{S}\subset\{1,\ldots, k-1\} } } \!\! (-1)^{|S|} \,
       \prod_{\substack{ j=1 \\ j\not\in\tilde{S}}}^{k-1} \Bigg\{
  \frac{a(v_{j})}{d(v_{j}) }\, \prod_{l=1}^n [u_{l}-v_{j}+1] \Bigg\}
   \\
   \times
    \prod_{j\in\tilde{S}} \Bigg\{(-1)^{r\aleph}\omega_u^{-2} \prod_{l=1}^n [u_{l}-v_{j}-1] \Bigg\}
    \det_n \big[ \check{\mathcal{N}}^{(k)}_\gamma (\{u\};v_{1}^{S\,\tilde{S} },\ldots, v_{{k-1}}^{S\,\tilde{S} }, \xi_{\ell_q}-1,\xi_{\ell_{k+1}}^S,\ldots,\xi_{\ell_n}^S  ) \big] 
   \displaybreak[0] \\
 -  \sum_{q=1}^{k-1} \frac{[s+N-2n+v_{k}-v_{q}+k] }{  [ |u| -|\xi_{\ell}|_{n-k}-|v|_{k-1}-v_{q}+\gamma+s] }
   \frac{\prod_{j=1}^N [v_{k}-\xi_j]}{[v_{k}-v_{q}]}
    \prod_{\substack{l=1 \\ l\ne q}}^{k-1}   \frac{1}{[v_{q}-v_{l}]}
    \prod_{j=k+1}^n\frac{1}{[\xi_{\ell_j}-v_{q}-1]} \\
    \times    
      \sum_{\substack{S\subset\{1,\ldots, n\}\setminus\{k\} \\ \tilde{S}\subset\{1,\ldots, k-1\} } }
       \prod_{\substack{ j=1 \\ j\not\in\tilde{S}}}^{k-1} \Bigg\{
  \frac{a(v_{j})}{d(v_{j}) }\, \prod_{l=1}^n [u_{l}-v_{j}+1] \Bigg\}
   \prod_{j\in\tilde{S}} \Bigg\{(-1)^{r\aleph}\omega_u^{-2} \prod_{l=1}^n [u_{l}-v_{j}-1] \Bigg\} \\
   \times (-1)^{|S|+|\tilde{S}|} \, \Bigg\{  \prod_{j=1}^N \frac{[v_{q}-\xi_j+1]}{[v_{q}-\xi_j]}\,
         \det_n \big[ \hat{\mathcal{N}}^{(k)}_\gamma (\{u\};v_{1}^{S\,\tilde{S} },\ldots, v_{k-1}^{S\,\tilde{S} }, v_{q},\xi_{\ell_{k+1}}^S,\ldots,\xi_{\ell_n}^S  ) \big] 
   \\
   -(-1)^{r\aleph}\omega_u^{-2} \det_n \big[ \check{\mathcal{N}}^{(k)}_\gamma (\{u\};v_{1}^{S\,\tilde{S} },\ldots, v_{k-1}^{S\,\tilde{S} }, v_{q},\xi_{\ell_{k+1}}^S,\ldots,\xi_{\ell_n}^S  ) \big]   
   \Bigg\} ,
         \label{fpter}
\end{multline}
with $\big[ \check{\mathcal{N}}^{(k)}_\gamma (\{u\};v_{1}^{S\,\tilde{S} },\ldots, v_{k-1}^{S\,\tilde{S} }, v_{k},\xi_{\ell_{k+1}}^S,\ldots,\xi_{\ell_n}^S  ) \big] _{ij}
                =\big[ {\mathcal{N}}_\gamma(\{u\};v_{1}^{S\,\tilde{S} },\ldots, v_{k-1}^{S\,\tilde{S} }, \xi_{\ell_k}^S,\ldots,\xi_{\ell_n}^S  ) \big]_{ij}$ for $j\ne k$ and 
\begin{multline}
   \big[ \check{\mathcal{N}}^{(k)}_\gamma (\{u\};v_{1}^{S\,\tilde{S} },\ldots, v_{k-1}^{S\,\tilde{S} },  v_{k},\xi_{\ell_{k+1}}^S,\ldots,\xi_{\ell_n}^S  ) \big]_{ik}
    \\
     =-\prod_{\substack{l=1\\ l\ne i} }^n[u_l-v_k-1]   \,  
        \frac{[1] [\gamma]  [u_i-v_k+s-|S|+|\tilde{S}|+\gamma]
                                         [  u_i-v_k-s+|S|-|\tilde{S}|  -1]   }
               {[s-|S|+|\tilde{S}| ] [s-|S|+|\tilde{S}|+1] [u_i-v_k] } 
    \displaybreak[0] \\
     = -\prod_{l=1}^n[u_l-v_k-1] \,
        \bigg\{\frac{[\gamma+s-|S|+|\tilde{S}| ]}{[s-|S|+|\tilde{S}| ]}
              \frac{[u_i-v_k+\gamma]}{[u_i-v_k]}  \\
          -     \frac{[\gamma+s-|S|+|\tilde{S}| +1]}{[s-|S|+|\tilde{S}|+1 ]}
              \frac{[u_i-v_k+\gamma-1]}{[u_i-v_k-1]} \bigg\}, 
\end{multline}
in which we have used \eqref{id1}.
Note that, due to the fact that $\{u\}$ satisfies the Bethe equations, there is no pole at $v_k=u_i$ in \eqref{fpter}.
Note also that, due to Remark~\eqref{rem-pole-part}, there is no pole at $|u| -|\xi_{\ell}|_{n-k}-|v|_k+\gamma+s=0$ since the corresponding sums over determinants have a zero at this point.
Therefore the function \eqref{fpter} is an entire function of the form \eqref{ratio-theta} with $m=N$ and $t=\sum_{j=1}^N\xi_j-s-N+2n-k$,
and it takes the same value than \eqref{fpbis} at the $N$ independent points $v_k=\xi_j$, $j=1,\ldots,N$.
Hence the equality between \eqref{fpbis} and \eqref{fpter}.

Up to the product $\prod_{j=1}^N[v_k-\xi_j]$, the second term in \eqref{fpter} is equal to
\begin{multline*}
   (-1)^{r\aleph}\omega_u^{-2}
   \sum_{q=k+1}^n \frac{[s+N-2n+v_k-\xi_{\ell_q}+k+1] }{ [ |u| -|\xi_{\ell}|_{n-k}-|v|_{k-1}-\xi_{\ell_q}+\gamma+s+1] }
   \frac{\prod_{l=1}^n [u_l-\xi_{\ell_q}]}{[\xi_{\ell_q}-v_k-1]}
    \prod_{\substack{l=k+1 \\ l\ne q}}^n \!  \frac{1}{[\xi_{\ell_l}-\xi_{\ell_q}]} \\
    \times
    \prod_{j=1}^{k-1}\! \frac{1}{[\xi_{\ell_q}-v_{j}-1]}
      \sum_{ \tilde{S}\subset\{1,\ldots, k-1\}  }
       \prod_{\substack{ j=1 \\ j\not\in\tilde{S}}}^{k-1}\! \Bigg\{\!
  \frac{a(v_{j})}{d(v_{j}) } \prod_{l=1}^n [u_{l}-v_{j}+1] \! \Bigg\}
   \prod_{j\in\tilde{S}}\! \Bigg\{\! (-1)^{r\aleph}\omega_u^{-2}
                                                  \prod_{l=1}^n [u_l-v_{j}-1] \! \Bigg\} \\
   \times \!\!
    \sum_{S\subset \{1,\ldots,n\}\cup\{ q\} } \hspace{-6mm} (-1)^{|S|} \,\frac{[\gamma+s-|S|+|\tilde{S}|+1]}{[s-|S|+|\tilde{S}|+1]}
    \det_n \! \big[ \mathcal{N}_\gamma (\{u\};v_{1}^{S\,\tilde{S} },\ldots, v_{k-1}^{S\,\tilde{S} }, \xi_{\ell_q}^S,\xi_{\ell_{k+1}}^S,\ldots,\xi_{\ell_n}^S  ) \big] ,
\end{multline*}
which is equal to zero since the sum over $S$ is antisymmetric with respect to the variables $\xi_{\ell}$, and since the variable $\xi_{\ell_q}$ appears twice in this sum.
Also, up to the product $\prod_{j=1}^N[v_{k}-\xi_j]$, the third term in \eqref{fpter} is equal to
\begin{multline*}\label{terme-zero}
 -\sum_{q=1}^{k-1} \frac{[s+N-2n+v_{k}-v_{q}+k] }{  [ |u| -|\xi_{\ell}|_{n-k}-|v|_{k-1}-v_{q}+\gamma+s] }
   \frac{1}{[v_{k}-v_{q}]}
    \prod_{\substack{l=1 \\ l\ne q}}^{k-1}   \frac{1}{[v_{q}-v_{l}]} 
    \prod_{j=k+1}^n\frac{1}{[\xi_{\ell_j}-v_{q}-1]}\\
   \times \!   
      \sum_{\substack{S\subset\{1,\ldots, n\} \cup\{q\}\setminus\{k\}\\ \tilde{S}\subset\{1,\ldots, k-1\}\cup\{q\} } } \,
       \prod_{\substack{ j\in\{1,\ldots, k-1\} \\ \cup\{q\} \setminus \tilde{S}}}\!\! \Bigg\{ 
  \frac{a(v_{j})}{d(v_{j}) } \prod_{l=1}^n [u_l-v_{j}+1] \Bigg\}
   \prod_{j\in\tilde{S}}\! \Bigg\{(-1)^{r\aleph}\omega_u^{-2} \prod_{l=1}^n [u_{l}-v_{j}-1] \Bigg\} \\
   \times  (-1)^{|S|+|\tilde{S}|}\frac{[\gamma+s-|S|+|\tilde{S}| ]}{[s-|S|+|\tilde{S}| ]}\,
         \det_n \big[ \mathcal{N}_\gamma (\{u\};v_{1}^{S\,\tilde{S} },\ldots, v_{k-1}^{S\,\tilde{S} },
          v_{q}^{S\,\tilde{S} },
                \xi_{\ell_{k+1}}^S,\ldots,\xi_{\ell_n}^S  ) \big]  ,
\end{multline*}
which is equal to zero due to the antisymmetry of the sum over determinants by exchange of two $v_{\beta}$ (the variable $v_q$ appears twice in this sum).
In the two above expressions, the notation $\{1,\ldots, n\} \cup\{q\}$ or $\{1,\ldots, k-1\} \cup\{q\}$ means a set that contains twice the element~$q$.

Finally, we obtain,
\begin{multline}
 f^{(k)}(v_{k})=\frac{[s+N-2n+k] }{ [ |u| -|\xi_{\ell}|_{n-k}-|v|_k+\gamma+s] }
 \frac{\prod_{j=1}^N[v_{k}-\xi_j]}{\prod_{j=k+1}^n [\xi_{\ell_j}-v_{k}-1] \cdot \prod_{j=1}^{k-1}[v_{k}-v_{j}]}
  \\
 \times \sum_{\substack{S\subset\{1,\ldots, n\} \\ \tilde{S}\subset\{1,\ldots, k\} } } (-1)^{|S|+|\tilde{S}|} \,
       \prod_{\substack{j=1 \\ j\not\in\tilde{S}}}^k \Bigg\{
  \frac{a(v_{j})}{d(v_{j}) }\, \prod_{l=1}^n [u_{l}-v_{j}+1] \Bigg\}
   \prod_{j\in\tilde{S}} \Bigg\{(-1)^{r\aleph}\omega_u^{-2} \prod_{l=1}^n [u_l-v_j-1] \Bigg\}
   \\
      \times  \frac{[\gamma+s-|S|+|\tilde{S}| ]}{[s-|S|+|\tilde{S}| ]}\,
         \det_n \big[ \mathcal{N}_\gamma (\{u\};v_{1}^{S\,\tilde{S} },\ldots, v_{k-1}^{S\,\tilde{S} },
          v_{k}^{S\,\tilde{S} },
                \xi_{\ell_{k+1}}^S,\ldots,\xi_{\ell_n}^S  ) \big]  ,
\end{multline}
which means that $G^{(k)}_{\ell_{k+1},\ldots, \ell_n} ( \{u\}; \{v_{1}, \ldots, v_{k} \} ; s)$ is given by \eqref{rec-Gk}.

\section{Orthogonality of two different Bethe eigenstates}
\label{app-orth}

We show in this Appendix that, similarly as what happens for the XXZ chain  \cite{KitMST05b}, the matrix $\Omega_\gamma(\{u\},\omega_u;\{v\},\omega_v)$ \eqref{def-Omega} giving the scalar product of two different Bethe eigenstates $\bra{ \{u\},\omega_u }$ and $\ket{\{v\},\omega_v }$, with $\{ u\}$ and $\{ v \}$ two different off-diagonal solutions of the Bethe equations \eqref{Bethe}, admits a zero eigenvalue.

Let us first suppose that $u_j\not= v_k, \forall j,k$.
Then, using the Bethe equations for $\{v\}$ and factorizing the product $ (-1)^{r\aleph} a(v_j)\prod_{l=1}^n [u_l-v_j+1]$ out of each column $j$ of the matrix $\Omega_\gamma (\{u\},\omega_u;\{v\},\omega_v)$, we obtain a matrix $M_\gamma(\{u\},\omega_u;\{v\},\omega_v)$ of elements
\begin{multline} 
   \big[ M_\gamma(\{u\},\omega_u;\{v\},\omega_v)\big]_{ij}
   =  \frac{1}{[\gamma]} \left\{ \frac{[u_i - v_j + \gamma]}{[u_i -v_j ]}
                    - \frac{\omega_v}{\omega_u} \frac{[u_i-v_j+\gamma +1]}{[u_i-v_j +1]} \right\}
    \\
      - \frac{1}{[\gamma]}\left\{ \frac{[u_i - v_j + \gamma]}{[u_i -v_j ]} 
                   - \frac{\omega_u}{\omega_v} \frac{[u_i-v_j+\gamma -1]}{[u_i-v_j -1]} \right\}
      \frac{\omega_v^2}{\omega_u^2}
        \prod_{l=1}^n \frac{[v_l - v_j + 1]}{[v_l - v_j- 1]} \frac{[u_l- v_j-1]}{[u_l- v_j + 1]} .
\end{multline}
The action of the transpose of the above matrix on the non-zero vector $\mathbf{w}$ of components
\begin{equation}
  \mathbf{w}_i=\frac{\prod_{l=1}^n [u_i-v_l]}{\prod_{\substack{l=1\\ l\not= i}}^n[u_i-u_l]}
\end{equation}
produces a new vector $\tilde{\mathbf{w}}$ of components
\begin{align}
   \tilde{\mathbf{w}}_i
   &=\sum_{j=1}^n \big[ M_\gamma(\{u\},\omega_u;\{v\},\omega_v)^t \big]_{ij}\, \mathbf{w}_j \nonumber\\
   &= \left\{ G_0-\frac{\omega_v}{\omega_u} G_1\right\}- \left\{ G_0- \frac{\omega_u}{\omega_v} G_{-1} \right\}
      \frac{\omega_v^2}{\omega_u^2}
        \prod_{l=1}^n \frac{[v_l - v_i + 1]}{[v_l - v_i- 1]} \frac{[u_l- v_i-1]}{[u_l- v_i + 1]} , \label{wtilde}
\end{align}
with
\begin{align}
  G_\eps
  &= \sum_{j=1}^n \frac{[u_j-v_i+\gamma+\eps]}{[\gamma]\, [u_j-v_i+\eps]} \frac{\prod_{l=1}^n [u_j-v_l]}{\prod_{\substack{l=1\\ l\not= j}}^n[u_j-u_l]}\label{ell1}
  \\
  &= -\prod_{l=1}^n \frac{[v_i-v_l-\eps]}{[v_i-u_l-\eps]}.\label{ell2}
\end{align}
The equality between \eqref{ell1} and \eqref{ell2} follows from the residue theorem applied to the elliptic function
\begin{equation}\label{feps}
  f_\eps(z)= \frac{[z-v_i+\gamma+\eps]}{[z-v_i+\eps]} \prod_{l=1}^n\frac{[z-v_l]}{[z-u_l]}
\end{equation}
integrated on an elementary cell (we recall here that $\gamma=|v|-|u|$, which means that the function $f_\eps$ \eqref{feps} is effectively doubly periodic, 
and that its integral on an elementary cell vanishes). 
Replacing the value \eqref{ell2} of $G_\eps$ into the expression \eqref{wtilde}, we get that $\tilde{\mathbf{w}}=0$, which means that the non-zero vector $\mathbf{w}$ is an eigenvector of $M_\gamma^t$ with zero-eigenvalue, and hence that the scalar product $\moy{\{u\},\omega_u\, |\, \{v\},\omega_v }$ vanishes.  

In the case when some (not all) of the variables $u_j$ coincide with some of the variables $v_k$, the proof is similar but one has first to take the limit in the corresponding columns of the determinant.

\section{Completeness of Bethe eigenstates}
\label{app-completeness}

In the untwisted case and for a rational parameter $\eta = \frac{1}{L}$, the completeness of Bethe eigenstates (with arbitrary multiplier through the transformation $s\to s+L$) was studied by Felder, Tarasov and Varchenko in \cite{FelTV97}.
More precisely, there was shown the following result (Corollary~19, Theorem~21 of \cite{FelTV97}):

\begin{thm}
\cite{FelTV97}\label{th-FTV}
Let $N=2n$ and $\eta=1/L$ for some odd integer $L> n$.
For generic $\alpha$, $\{\xi\}$,  there are $d=L\dim\mathcal{H}[0]$ admissible\footnote{See footnote~\ref{foot-admissible}.} off-diagonal\footnote{See footnote~\ref{foot-odiag}.} solutions of the Bethe equations
\begin{equation}\label{betheTV}
  \prod_{k=1}^N\frac{[v_j-\xi_k+1]}{[v_j-\xi_k]}\prod_{\substack{l=1\\ l\neq j}}^n\frac{[v_j-v_l-1]}{[v_j-v_l+1]}=\omega^{-2},\qquad j=1,\ldots,n,
\end{equation}
with $\omega^L=(-1)^n\alpha$ such that the corresponding Bethe eigenstates \eqref{state}-\eqref{phi-FV} form a basis of the $d$-dimensional vector space of quasi $L$-periodic functions $f\in \mathrm{Fun}(\mathcal{H}[0])$ with multiplier $\alpha$, i.e. such that $f(s+L)=\alpha f(s)$.
\end{thm}

For our purpose, we need a slightly modified version of Theorem~\ref{th-FTV}:
\begin{thm}\label{th-compl}
Let $N=2n$ and $\eta=r/L$ for some relatively prime integers $L,r$ with $L$ odd and $L> n$.
There exist $\kappa_0>0$ such that, for $0<|\kappa|<\kappa_0$ and generic $\{\xi\}$,  there are $d=L\dim\mathcal{H}[0]$ admissible off-diagonal solutions of the $\kappa$-twisted Bethe equations
\begin{equation}\label{kappa-bethe}
  \prod_{k=1}^N\frac{[v_j-\xi_k+1]}{[v_j-\xi_k]}\prod_{\substack{l=1\\ l\neq j}}^n\frac{[v_j-v_l-1]}{[v_j-v_l+1]}=\kappa\,\omega^{-2},\qquad j=1,\ldots,n,
\end{equation}
with $\omega^L=(-1)^{rn}$, and the corresponding $\kappa$-twisted Bethe eigenstates \eqref{state}-\eqref{phi-FV} form a basis of the $d$-dimensional vector space of functions $f\in \mathrm{Fun}(\mathcal{H}[0])$ such that $f(s+L)= f(s)$.
\end{thm}

The proof of Theorem~\ref{th-FTV} in \cite{FelTV97} follows from the study the following eigenvalue problem with multiplier condition (this problem is related to the study of solutions of qKZB equations for the elliptic quantum group $E_{\tau,\eta}(sl_2)$):
\begin{align*}
    &H_j(\xi)\,\psi=\epsilon_j\,\psi, \qquad j=1,\ldots,N, \\
    &\psi\in\mathrm{Fun}(\mathcal{H}[0])\quad \text{such that}\quad \psi(s+L)=\alpha\psi(s).
\end{align*}
Here $H_j(\xi)$ are commuting difference operators defined in terms of $R$-matrices of the elliptic quantum group $E_{\tau,\eta}(sl_2)$:
\begin{equation}\label{diff-op}
  H_j(\xi)
  =R_{j,j-1}\ldots R_{j,1}\, \widehat{\tau}_s^{h_j}\, R_{j,N}\ldots R_{j,j+1},
\end{equation}
where $\xi=(\xi_1,\ldots,\xi_N)$ is a fixed generic point in $\mathbb{C}^N$, and where we have used the shorthand notation
\begin{equation}
   R_{j,k}\equiv R_{j,k}\big(\xi_j-\xi_k; \widehat{s}+\sum_{\substack{l=1\\ l\neq j}}^{k-1}h_l\big).
\end{equation}
In particular, in the fundamental case, the difference operators \eqref{diff-op} coincide with special values of the transfer matrix of the SOS model, and their eigenfunctions (for generic $\xi$) coincide with Bethe eigenstates.

The first part of the proof of \cite{FelTV97} consists in constructing common eigenfunctions to the operators $H_j(\xi)$, $j=1,\ldots,n$. To this aim, the spaces dual to tensor products of evaluation Verma modules over $E_{\tau,\eta}(sl_2)$ are realized as particular spaces of meromorphic functions of $n+1$ complex variables (typically the $n$ spectral parameters and the dynamical parameter) with adequate quasi-periodicity properties. A basis of this space of functions is explicitly identified.
Common eigenfunctions (with given multiplier $\alpha$) of the commuting difference operators $H_j(\xi)$ are then constructed in terms of these basis elements evaluated at the solutions of the system of Bethe equations \eqref{betheTV}. These eigenfunctions are identified (up to a factor that vanishes for diagonal solutions of \eqref{betheTV}) with Bethe eigenstates.
We refer the reader to \cite{FelTV97} for explicit details about this construction, which is valid for any value of $\eta$ and of the multiplier $\alpha$.

The second part of the proof of \cite{FelTV97} consists in showing that,  in the case $n=N/2$, $\eta=1/L$, and for generic $\alpha$, the previously constructed set of eigenfunctions is indeed complete.
For $\alpha$ large but finite, one can identify $L\dim  \mathcal{H}[0]$ different
sets of admissible off-diagonal solutions of the system \eqref{betheTV}.
Using the explicit expressions, previously obtained through the aforementioned construction, of the corresponding eigenfunctions,  the authors of \cite{FelTV97} are finally able to show that the latter are indeed linearly independent, at least for $\alpha$ large enough
(see \cite{FelTV97} for details).

To prove Theorem~\ref{th-compl}, one needs to extend these arguments to the $\kappa$-twisted case with multiplier 1. It is in fact easy to see that, in the previous reasoning, one can use $\kappa$ as a free parameter instead of $\alpha$. We use the fact that solutions of the system of Bethe equations at $\kappa=0$ are very simple and that, for $L$ odd, there exist $L$ possible distinct values of the parameter $\omega^2$ such that $\omega^L=(-1)^{rn}$.
Hence, considering the system \eqref{kappa-bethe} when $\kappa$ tends to zero and using the implicit function Theorem, one can identify $L\dim  \mathcal{H}[0]$ different\footnote{up to periodicity in $1/\eta$ and $\tau/\eta$. The condition $L$ odd and $L>n$ ensures that  the zero weight space $\mathcal{H}[0]$ reduces to the space corresponding to $n=N/2$.} sets of admissible off-diagonal solutions $\{v^{I,\ell}_j(\kappa)\}_{1\le j\le n}$ (corresponding to all possible subsets $I=(i_1,\ldots,i_n)\subset\{1,\ldots,N\}$ and all possible integers $\ell\in\{0,\ldots L-1\}$) which are continuous functions of $\kappa$ in a vicinity of $\kappa=0$. They are such that
\begin{equation}\label{sol-0}
   v^{I,\ell}_j(\kappa)
   = \xi_{i_j}-1+\kappa\, e^{4\pi i \frac{\ell}{L}}\, u_j^{I}(0)+o(\kappa), 
\end{equation}
with
\begin{equation}\label{uj-0}
   u_j^{I}(0)=  e^{2i\pi\frac{rn}{L}}\frac{[1]}{[0]'} \prod_{\substack{k=1 \\ k\not= i_j}}^N \frac{[\xi_{i_j}-\xi_k-1]}{[\xi_{i_j}-\xi_k]}
   \prod_{k=1}^n\frac{[\xi_{i_k}-\xi_{i_j}-1]}{[\xi_{i_k}-\xi_{i_j}+1]}.
\end{equation}

To show that the corresponding Bethe eigenstate form a basis of the corresponding space of functions, i.e. that they are linearly independent, one can adapt the arguments of \cite{FelTV97} to the $\kappa$-twisted case by explicitly constructing common eigenfunctions to the $\kappa$-twisted commuting difference operators
\begin{equation}
  H_j^{(\kappa)}(\xi)
  =R_{j,j-1}\ldots R_{j,1}\, \widehat{\tau}_s^{h_j}\, K_j^{(\kappa)}\, R_{j,N}\ldots R_{j,j+1},
  \qquad \text{with}\quad K^{(\kappa)}=\begin{pmatrix} 1 & 0\\ 0 & \kappa \end{pmatrix}.
\end{equation}
The linear independence, for $\kappa$ small enough, of the eigenfunctions  corresponding to solutions \eqref{sol-0} then follows from the same reasoning as in \cite{FelTV97}, which can easily be extended to the case of rational $\eta=r/L$.

Alternatively, one can remark that:
\begin{itemize}
  \item the scalar product \eqref{sc-kappa} of two Bethe states corresponding to different off-diagonal solutions of the $\kappa$-twisted Bethe equations \eqref{kappa-bethe} vanishes (the proof of Appendix~\ref{app-orth} is indeed directly generalizable to the $\kappa$-twisted case).
  \item the ``square of the norm'' \eqref{norm-kappa} of a $\kappa$-twisted Bethe state corresponding to an admissible off-diagonal solution of \eqref{kappa-bethe} is non-zero in a vicinity of $\kappa=0$.
Namely,
one has,
\begin{equation*}
   \frac{\partial}{\partial v_k} \mathcal{Y}_{\kappa;\omega}(v_j |\{v\}_{_{\!\kappa}})
   =\kappa^{-1} \Bigg\{ \omega^2 \delta_{j,k} \frac{[1]}{[0]'\, u_j(0)^2}
     \prod_{\substack{k=1 \\ k\not= i_j}}^N \frac{[\xi_{i_j}-\xi_k-1]}{[\xi_{i_j}-\xi_k]} \prod_{k=1}^n{[\xi_{i_k}-\xi_{i_j}-1]} + o(\kappa)\Bigg\},
\end{equation*}
for $\{v\}\equiv\{v^{I,\ell}\}$ given by \eqref{sol-0} with $\omega=e^{i\pi\frac{rn}{L}-2i\pi \frac{\ell}L}$ and $u_j(0)\equiv u_j^I(0)$ given by \eqref{uj-0}.
\end{itemize}
This prove the linear independence, for $0<|\kappa|<\kappa_0$ for some $\kappa_0>0$, of the $L\dim\mathcal{H}[0]$ $\kappa$-twisted Bethe eigenvectors corresponding to the solutions \eqref{sol-0}, and hence ends the proof of Theorem~\ref{th-compl}.



\providecommand{\bysame}{\leavevmode\hbox to3em{\hrulefill}\thinspace}
\providecommand{\MR}{\relax\ifhmode\unskip\space\fi MR }
\providecommand{\MRhref}[2]{%
  \href{http://www.ams.org/mathscinet-getitem?mr=#1}{#2}
}
\providecommand{\href}[2]{#2}

\end{document}